\documentclass{aa}  

\usepackage{graphicx}
\usepackage{txfonts, natbib}
\usepackage{hyperref}
\usepackage{color}

\begin{document}

\title{Radial velocity observations of the 2015 Mar 20 eclipse}
\subtitle{A benchmark Rossiter-McLaughlin curve with zero free parameters}

\author{A. Reiners\inst{1}
  \and
  U. Lemke\inst{1}
  \and 
  F. Bauer\inst{1}
  \and
  B. Beeck\inst{2}
  \and
  P. Huke\inst{1}
}

\institute{Georg-August Universit\"at G\"ottingen, Institut f\"ur Astrophysik, 
  Friedrich-Hund-Platz 1, 37077 G\"ottingen, Germany\\
  \email{Ansgar.Reiners@phys.uni-goettingen.de}
  \and
  Max-Planck Institut f\"ur Sonnensystemforschung,
  Justus-von-Liebig-Weg 3, 37077 G\"ottingen, Germany
}

\date{\today}

 
\abstract{Spectroscopic observations of a solar eclipse can provide
  unique information for solar and exoplanet research; the huge
  amplitude of the Rossiter-McLaughlin (RM) effect during solar
  eclipse and the high precision of solar radial velocities (RVs)
  allow detailed comparison between observations and RV models, and
  they provide information about the solar surface and about spectral
  line formation that are otherwise difficult to obtain. On March 20,
  2015, we obtained 159 spectra of the Sun as a star with the solar
  telescope and the Fourier Transform Spectrograph at the Institut
  f\"ur Astrophysik G\"ottingen, 76 spectra were taken during partial
  solar eclipse. We obtained RVs using $I_2$ as wavelength reference
  and determined the RM curve with a peak-to-peak amplitude of almost
  1.4\,km\,s$^{-1}$ at typical RV precision better than
  1\,m\,s$^{-1}$. We modeled the disk-integrated solar RVs using
  well-determined parameterizations of solar surface velocities, limb
  darkening, and information about convective blueshift from 3D
  magnetohydrodynamic simulations. We confirm that convective
  blueshift is crucial to understand solar RVs during eclipse. Our
  best model reproduced the observations to within a relative
  precision of 10\,\% with residuals lower than 30\,m\,s$^{-1}$. We
  cross-checked parameterizations of velocity fields using a
  Dopplergram from the Solar Dynamics Observatory and conclude that
  disk-integration of the Dopplergram does not provide correct
  information about convective blueshift necessary for m\,s$^{-1}$ RV
  work. As main limitation for modeling RVs during eclipses, we
  identified limited knowledge about convective blueshift and line
  shape as functions of solar limb angle. We suspect that our model
  line profiles are too shallow at limb angles larger than $\mu = 0.6,$
  resulting in incorrect weighting of the velocities across the solar
  disk. Alternative explanations cannot be excluded, such as suppression
  of convection in magnetic areas and undiscovered systematics during
  eclipse observations. To make progress, accurate observations of
  solar line profiles across the solar disk are suggested. We publish
  our RVs taken during solar eclipse as a benchmark curve for codes
  calculating the RM effect and for models of solar surface velocities
  and line profiles.  }


   \maketitle
%

\section{Introduction}

The observation of radial velocities (RVs) of the Sun as a star is gaining
interest in both solar and exoplanet communities. For solar applications, the
availability of methods for computing one single parameter (the
RV) from sometimes
complex solar data and its comparison to other proxies opens new possibilities
to search for physical relations. For the exoplanet community, the Sun is the
one benchmark object on which methods can be tested with precision
observations that offer more information than is available for any other star,
such as limb darkening and surface velocity fields.

Obtaining RVs in stars involves measuring the Doppler shift of
spectral lines at different times of observation. There are several
ways to precisely determine a shift of spectral lines. They all have
in common that a change in the line profile shape caused by
an emerging active region on the star, for instance, can mimic a Doppler shift of
several m\,s$^{-1}$\citep[see, e.g.,][and references
therein]{2016PASP..128f6001F}. Finding ways to account for this type
of bias is currently an important step in the search for Earth-like
extrasolar planets \citep[e.g.,][]{2013ApJ...770..133H,
  2014Sci...345..440R, 2015ApJ...805L..22R, 2015arXiv150609072A}. In
the context of stars, it has been realized that m\,s$^{-1}$ precision
can only be reached when stellar surface velocities, and in particular
the effect of magnetic activity on convective blueshift, are
thoroughly understood \citep{2010A&A...512A..38L, 2010A&A...512A..39M,
  2010A&A...519A..66M, 2014ApJ...796..132D, 2015ApJ...798...63M,
  2016MNRAS.457.3637H}.

The Sun provides an invaluable reference for understanding the
influence of stellar surface velocity fields and active regions
because it is possible to relate precision RV observations to
spatially resolved solar surface information
\citep{2010A&A...519A..66M, 2016MNRAS.457.3637H}. However, it is
extremely difficult to obtain solar disk observations with m\,s$^{-1}$
precision or better, and current solar Dopplergrams are known to
exhibit severe problems in this respect \citep{2012SoPh..275..285C,
  2013ApJ...765...98W}. Furthermore, measuring precise Sun-as-a-star
RVs is very challenging because the spatial extension of the Sun
implies that feeding light from the Sun into a spectrograph involves
the problem of collecting light from all areas of the solar disk
equally. Earlier attempts have demonstrated this difficulty
\citep{Jimenez1986AdSpR, Deming1987ApJ, Deming1994ApJ,
  McMillan1993ApJ} but there is new motivation for instruments facing
this challenge, for example, observing the Sun with HARPS
\citep{2015ApJ...814L..21D} and in the G\"ottingen Solar Radial
Velocity Project used for this work \citep{2016arXiv160300470L}.

A particularly fruitful exercise to understand the effects across an
extended stellar disk in RV measurements is the observation of an
eclipse of the Sun by one of its planets or the Moon. During eclipse,
solar and stellar RVs exhibit the so-called Rossiter-McLaughlin (RM)
effect \citep{1924ApJ....60...15R, 1924ApJ....60...22M} with excursion
of the RV curve to the blue and the red in proportion to the
rotational velocity of the star and to the amount of the surface
eclipsed \citep[e.g.,][]{2005ApJ...622.1118O, 2007ApJ...655..550G,
  2008ApJ...677.1324T, 2009ApJ...696.1230F}. Observations of planetary
transits of the Sun can also be performed with night-time telescopes
observing the light reflected from other bodies of the solar system in
order to avoid the effect of the spatially extended solar disk
\citep[e.g.,][]{2013MNRAS.429L..79M, 2015MNRAS.453.1684M}. The RM
effect in stars is used to determine the geometry of planetary
systems. Convective blueshift influences the shape of the RM curve and
potentially biases measurements of planetary system geometries
\citep{2011ApJ...733...30S, 2013A&A...550A..53B, 2016ApJ...819...67C,
  2016A&A...588A.127C}. A planetary transit can also be used to
observe local stellar line profiles \citep{2015csss...18..853D}, and
in the same sense, observations of the eclipsed Sun provide a unique
opportunity to gather information about local solar line profiles and about the
influence of convective blueshift on RM measurements.

\citet{2015PASJ...67...10T} carried out observations of the solar
eclipse visible in Japan on May 21, 2012, with the aim to extract
information on solar rotation. At maximum, the Moon eclipsed 93\,\% of
the visible Sun during their observations. They showed an RM curve with
a peak-to-peak amplitude of almost 2\,km\,s$^{-1}$ and calculated a
model of the RV curve that matches within a standard deviation of
41\,m\,s$^{-1}$. Their model includes solar differential rotation and
limb darkening, and they attempted to gain information on solar surface
rotation from their RV curve. \citet{2015PASJ...67...10T} also provided
information about the spectral line shape during eclipse and discussed
deviations between ATLAS9 solar model calculations
\citep{1993KurCD..13.....K} and their observations. 

In this work, we investigate our solar eclipse observations from Mar
15, 2015, with a maximum eclipse of 75.9\,\%. We argue that solar
surface rotation and large-scale granulation patterns are relatively
well described from different tracers \citep{1992ASPC...27..205S,
  2004A&A...428.1007D}. On the other hand, the effect of convective
blueshift is less well quantitatively understood, in particular its
variation across the solar disk
\citep[e.g.,][]{1978SoPh...58..243B}. In addition, convective flows
can be suppressed in the presence of magnetic fields
\citep{1982Natur.297..208L}, with the consequence that magnetic areas
affect RV observations of active stars and the Sun. Therefore, we
assume in our work that the geometry of the eclipse as well as surface
rotation and granulation velocity fields are correctly known from
independent information. As the most critical parameters for modeling
solar RVs we identify line intensity changes across the solar disk and
the dependence of convective blueshift on solar limb angle.

\section{Instrument}
\label{sect:Instrument}

The Institut f\"ur Astrophysik, G\"ottingen (IAG) operates a Fourier
transform spectrograph (FTS) Bruker IFS 125 HR. Details of the
instrument and observing modes are described in
\cite{2016A&A...587A..65R} and \cite{2016arXiv160300470L}. We observed
the Sun as a star by feeding light of a 52' field of view from the
flat mirrors of a siderostat into a 800\,$\mu$m fiber that transports
the light into our optics laboratory. We sent the light through an
absorption cell filled with $I_2$ and into our FTS.  \cite{2016arXiv160300470L} showed that the photon noise limit in
our radial velocity (RV) observations is below 10\,cm\,s$^{-1}$ but
stellar and instrumental noise limit our RV precision; we found RV
drifts on the order of 10\,m\,s$^{-1}$\,h$^{-1}$ in other measuring
campaigns. Based on this experience, we cannot exclude drifts of this
amplitude for the observations used in this work. These drifts,
however, were typically rather linear for many exposures. In addition,
the RV curves sometimes showed sudden jumps of a few m\,s$^{-1}$ , but
they could be easily identified. They were probably caused by manual
readjustment of the siderostat tracking mechanism.

A second problem we only realized after observations is improper
logging of the start time of exposure. The FTS control software logged
the time of observation as given by the control computer. This machine
unfortunately was not properly synchronized to a time standard, with
the consequence that the absolute time of observation is unknown with
an uncertainty of several ten seconds, which is the typical time we
found the clock to be off.

\section{Observations}
\label{sect:Observations}

\begin{figure*}
  \centering
  \hspace{.023\hsize}
  \resizebox{.10\hsize}{!}{\includegraphics[bb =    0 1120 562.5 1680, clip=]{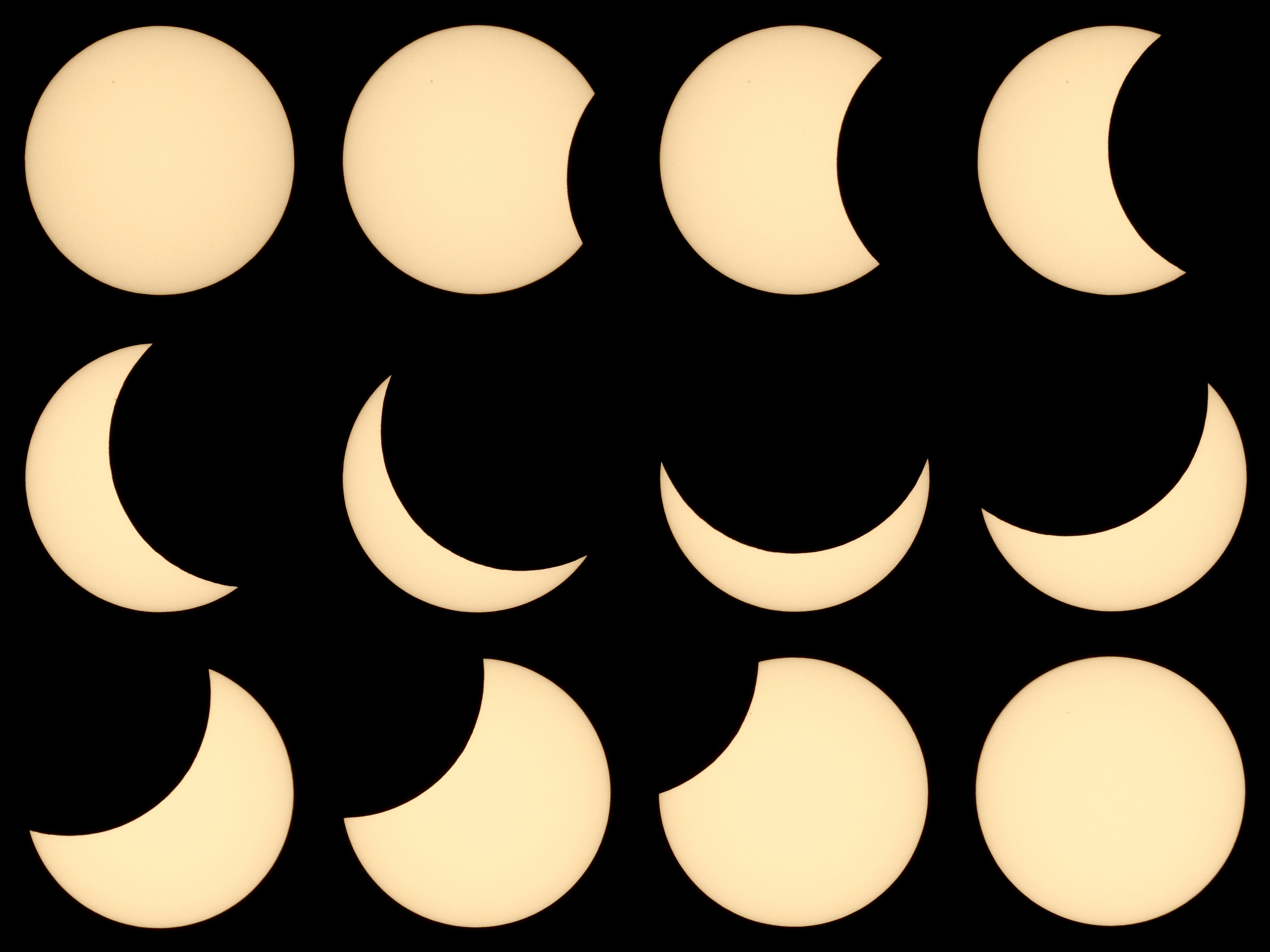}}\hspace{-1mm}
  \resizebox{.10\hsize}{!}{\includegraphics[bb = 562.5 1120 1125 1680, clip=]{eclipse_series.jpg}}\hspace{-1mm}
  \resizebox{.10\hsize}{!}{\includegraphics[bb =    0  560 562.5 1120, clip=]{eclipse_series.jpg}}\hspace{-1mm}
  \resizebox{.10\hsize}{!}{\includegraphics[bb = 1125 560 1687.5 1120, clip=]{eclipse_series.jpg}}\hspace{-1mm}
  \resizebox{.10\hsize}{!}{\includegraphics[bb =    0    0 562.5  560, clip=]{eclipse_series.jpg}}\hspace{-1mm}
  \resizebox{.10\hsize}{!}{\includegraphics[bb = 1125    0 1687.5 560, clip=]{eclipse_series.jpg}}\hspace{-1mm}
  \resizebox{.10\hsize}{!}{\includegraphics[bb = 1687.5  0 2250   560, clip=]{eclipse_series.jpg}}\\
  \resizebox{ .8\hsize}{!}{\includegraphics[bb =   10  100   640  450, clip=]{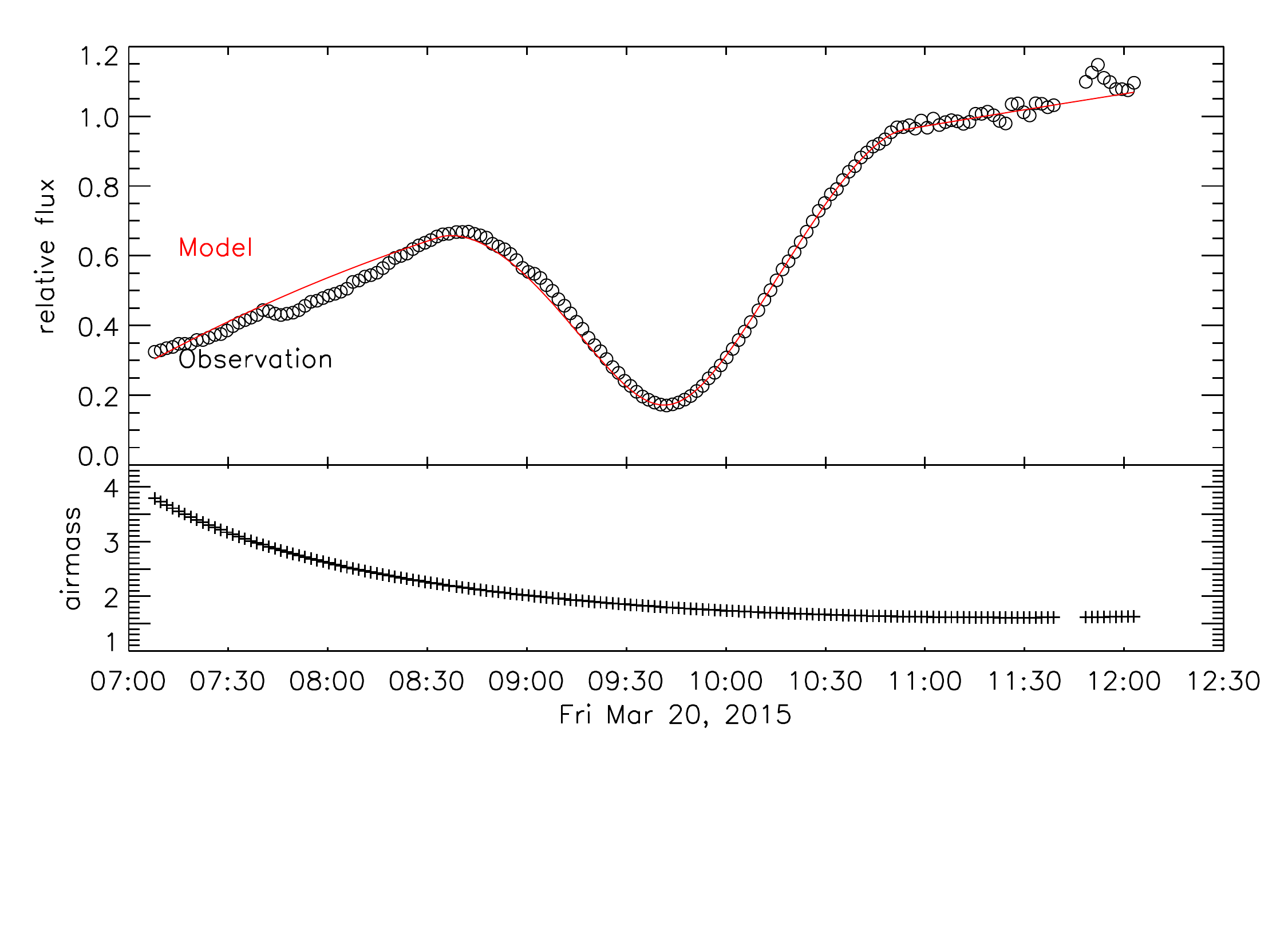}\hspace{2.8mm}}
  \caption{\label{fig:Flux}\emph{Upper panel:} Photographs of the
    solar eclipse taken at our observatory; \emph{middle panel:}
    observed flux normalized to 1.0 at 11:00\,UT, black open circles
    show flux integrated from our observed spectra, the red line shows
    the flux from our extinction model and eclipse (see text);
    \emph{bottom panel:} airmass during our observations. Times are
    given in UT.}
\end{figure*}

The solar eclipse occurred in G\"ottingen in the morning hours of
March 20, 2015, with its maximum at 09:42\,UT (10:42 local time). We
obtained 159 observations of the Sun as a star between 07:05 and
12:05\,UT. Details of observations are provided in
Table\,\ref{tab:Observations}. The interferograms were sampled in
fully symmetric forward-backward mode, each scan took 97\,s. The
resolution, $R$, of the transformed spectra is several 10$^5$. The
first spectrum was taken at airmass 3.8. At the start of the eclipse
(08:35\,UT), the Sun was at airmass 2.2. After the eclipse the Sun
culminated at airmass 1.6. Our spectra are sampled with constant step
size of 0.01\,cm$^{-1}$, that is, $\Delta \lambda = 0.0055$\,\AA\ at
6000\,\AA. We measured the signal-to-noise ratio (S/N) of our spectra
in the spectral range 6638.5--6641.1\,\AA. The typical S/N is higher
than 150 per sampled step, it increases to above 200 with decreasing
airmass and decreases to below 130 during the center of eclipse.

Figure\,\ref{fig:Flux} shows airmass and integrated flux in the
wavelength range 5500--6000\,\AA\ from our observations. In the top
panel we indicate phases of the eclipse with photographs of the Sun
taken by F.~Bauer. Our model of the eclipse is overplotted, showing the
effects of extinction and eclipse on the total flux. The model is
explained in detail in Sect.\,\ref{sect:Modeling}. The change of
airmass during our observations is shown in the bottom panel.

\section{Measuring radial velocities}
\label{sect:RVs}

To measure the RM curve from our spectra, we calculated the
RV from each individual spectrum relative to the barycentric corrected
solar atlas introduced in \citet{2016A&A...587A..65R}. The wavelength
scale of our instrument was monitored by simultaneously fitting the
absorption spectrum of $I_2$. We used the wavelength ranges $\lambda$
= 5000--6200\,\AA\ and $\lambda$ = 6500--6700\,\AA,\ excluding the
range contaminated by our internal HeNe laser.  For the fit, we
followed the recipes described in \citet{2012ApJS..200...15A}; free
parameters are the Doppler shift of $I_2$ and the Sun, and two
parameters of a linear continuum correction (offset and slope).

Radial velocities for all observations are reported in
Table\,\ref{tab:Observations}. Formal uncertainties from photon noise
are well below 1\,m\,s$^{-1}$ , but systematic differences between the
RVs are much larger because (i) our sampling rate is shorter than two
minutes and captures parts of the solar five-minute oscillations, (ii)
snapshot observations of convective motion on the Sun and variability
of active regions may lead to real RV jitter on the m\,s$^{-1}$ level,
and (iii) systematic offsets are introduced by our optical system
\citep[see][and Sect.\,\ref{sect:Results}]{2016arXiv160300470L}.

\section{Eclipse modeling}
\label{sect:Modeling}

\subsection{Solar and lunar ephemeris}

A solar eclipse is fully determined by our knowledge about the orbital
parameters of the Earth and the Moon. Parameters for the eclipse valid
for the location of our telescope\footnote{latitude: $51\degr\
  33\arcmin\ 38.1\arcsec$\ N; longitude: $009\degr\ 56\arcmin\
  39.6\arcsec$\ E;\\ altitude: 201\,m} were taken from JPL's HORIZONS
Web interface\footnote{http://ssd.jpl.nasa.gov/horizons.cgi}. For the
Sun and the Moon, we retrieved positions in right ascension and
declination, solar and lunar apparent radii, orientation of the solar
rotation axis, and projected velocities of the Sun. We assumed that
both bodies are perfect circles, which is an appropriate assumption
for our purpose \citep[for a discussion about the shape of the Moon,
see][]{iz2011polyaxial}.

\subsection{Limb darkening and velocity fields}
\label{sect:parameterization}

Several parameters are relevant for the calculation of disk-integrated
solar RVs.  Across the extended solar disk, projected velocities
differ by about $\pm 2$\,km\,s$^{-1}$ from one edge of the solar
equator to the other, velocities also differ between latitudes because
of surface differential rotation. Flux from individual regions is
weighted by limb darkening and also by cool sunspots. In addition,
spectral lines are Doppler shifted by convective motion that also
depends on the position on the solar disk. In
Sect.\,\ref{sect:Results} we show how different approaches to
parameterize velocities compare to our observations. Standard
descriptions for these mechanisms are the following:

\begin{enumerate}

\item{Limb darkening:} Solar limb darkening was determined for several
  wavelength regions by \citet{1994SoPh..153...91N}. We used the values
  determined for 580\,nm with $\mu$ the cosine of the solar zenith
  angle (solar limb angle) and $I$ the intensity:
  \begin{eqnarray}
    I(\mu) & = & 0.28392 + 1.36896\,\mu - 1.75998\,\mu^2 + 2.22154\,\mu^3 \nonumber \\ 
    && - 1.56074\,\mu^4 + 0.44630\,\mu^5
  .\end{eqnarray}

\item{Differential rotation:} We used the rotation law of the Sun as
  measured by \citet{1990ApJ...351..309S} with $l$ the solar latitude
  and $\varv$ the angular sidereal velocity in degrees per day
  \citep[see][for a discussion of solar differential rotation and of
  observational challenges of solar velocity field
  measurements]{1985SoPh..100..141S}:
  \begin{equation}
    \varv(l) = 14.713 - 2.396 \sin^2{l} - 1.787 \sin^4{l}
  .\end{equation}
  Because of the orbital motion of Earth, we converted from sidereal into
  synodic rotation velocity:
  \begin{equation}
    \varv_{\rm synodic} = 0.9324\,\varv_{\rm sidereal}
  .\end{equation}

  \begin{figure}
    \centering
    \resizebox{\hsize}{!}{\includegraphics[bb = 10 20 620 290]{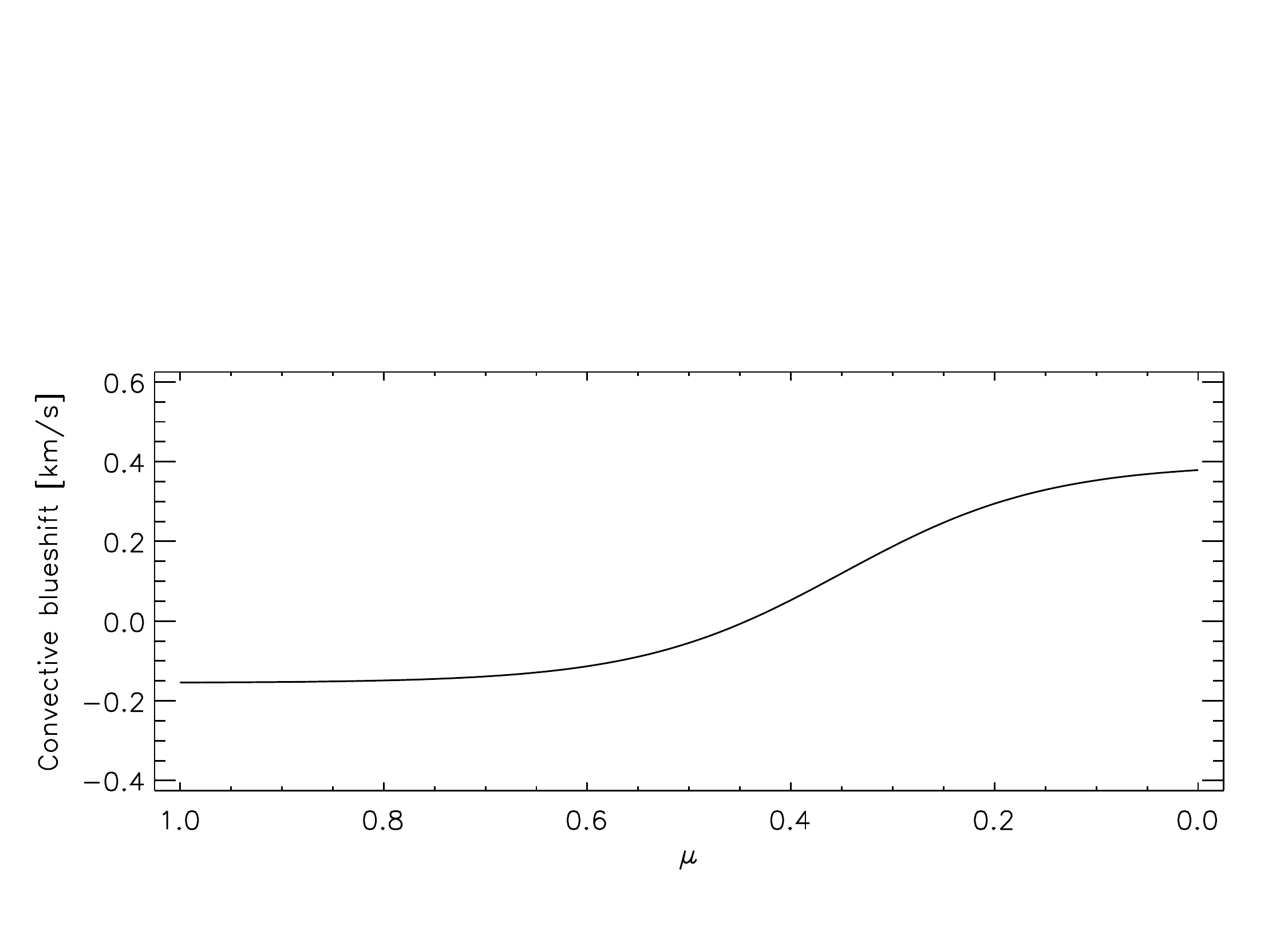}}
    \caption{\label{fig:Blueshift}Parameterization of convective blueshift as a
      function of solar limb angle. The curve qualitatively reproduces results
      from 3D magneto-hydrodynamic simulations and provides a useful fit to
      the observed RV curve; see text for details.}
  \end{figure}

\item{Convective blueshift:} Granular motion introduces velocity
  fluctuations of the photosphere plasma in horizontal and vertical
  direction. The correlation between velocity and temperature
  (intensity) produces systematic Doppler shifts that vary over the
  solar disk \citep{1978SoPh...58..243B, 2009ApJ...697.1032G}. In
  general, the effect is quite well understood and can be modeled in
  (magneto-) hydrodynamic simulations
  \citep[e.g.,][]{2013A&A...558A..49B}. However, the blueshift can vary
  between different spectral lines and sensitively depends on solar
  limb angle and the presence of magnetic fields because convection is
  suppressed in sunspots and magnetically active regions. Reference
  observations of the effect require very high wavelength accuracy and
  knowledge about the position on the solar surface to correct for
  Doppler shifts from solar rotation; such observations are available
  for one position 10\,\arcsec\ from the solar limb in
  \citet{2015A&A...573A..74S}, but were measured at limited wavelength
  accuracy \citep[see][]{2016A&A...587A..65R}. We parameterized
  the convective blueshift as a function of the solar limb angle $\mu$
  using a sigmoid function. The parameters of the function are chosen
  to qualitatively follow convective blueshift as shown in Fig.\,18 of
  \citet{2013A&A...558A..49B} for the Sun, and to provide a reasonable
  fit in the RM curve. Our parameterization is shown in
  Fig.\,\ref{fig:Blueshift}, it is neither an accurate reproduction of
  any model simulation nor the result of a systematic fit to the
  observed RVs. The purpose of this parameterization is to demonstrate
  the influence of convective blueshift on the RM curve. Nevertheless,
  we extensively tried to find parameterizations that provide a better
  fit to the RV curve, but with limited success.

\end{enumerate}

\begin{figure*}
  \centering
  \mbox{
    \resizebox{.33\hsize}{!}{\includegraphics[bb = 107 12 540 445]{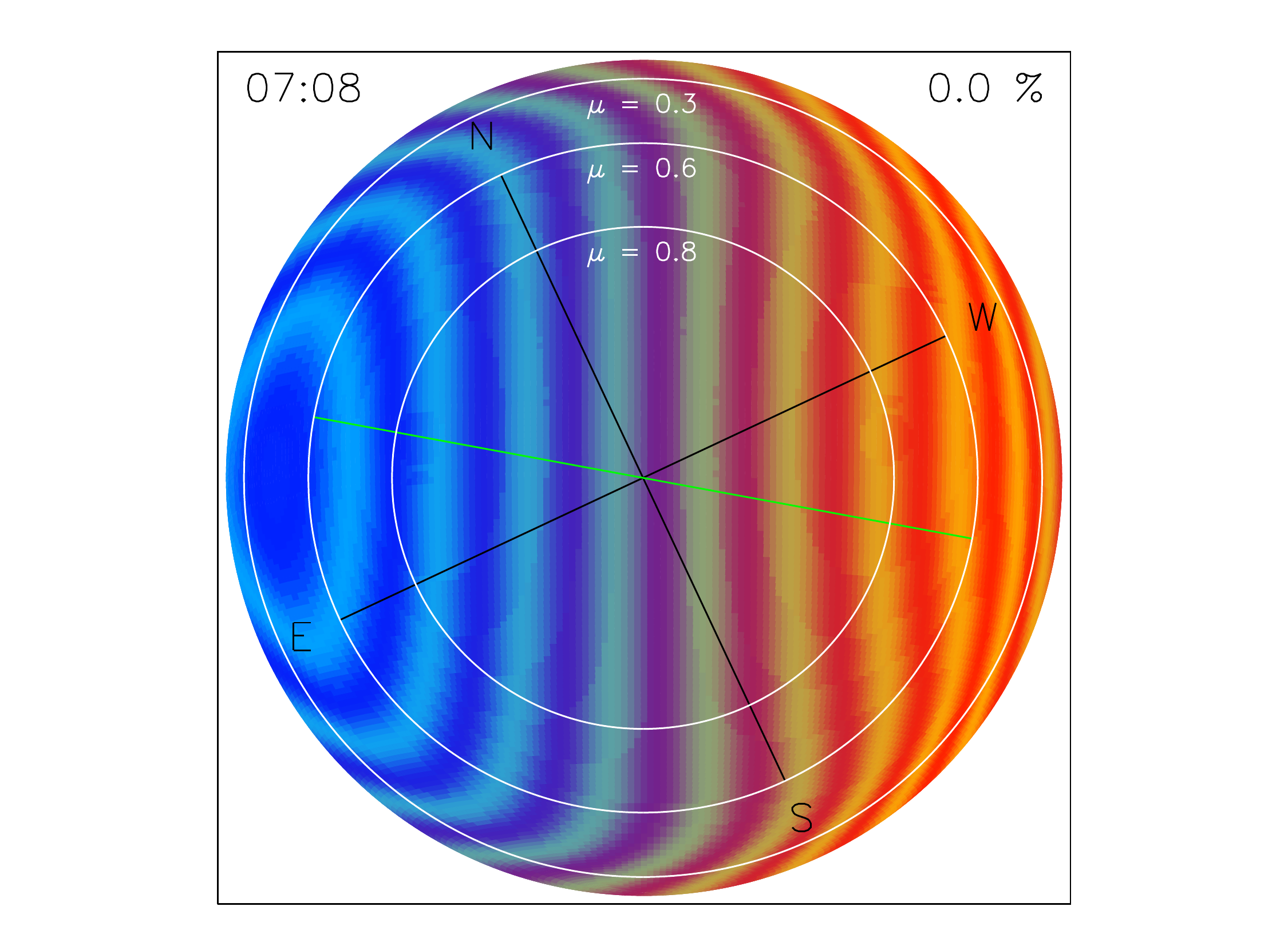}}
    \resizebox{.33\hsize}{!}{\includegraphics[bb = 107 12 540 445]{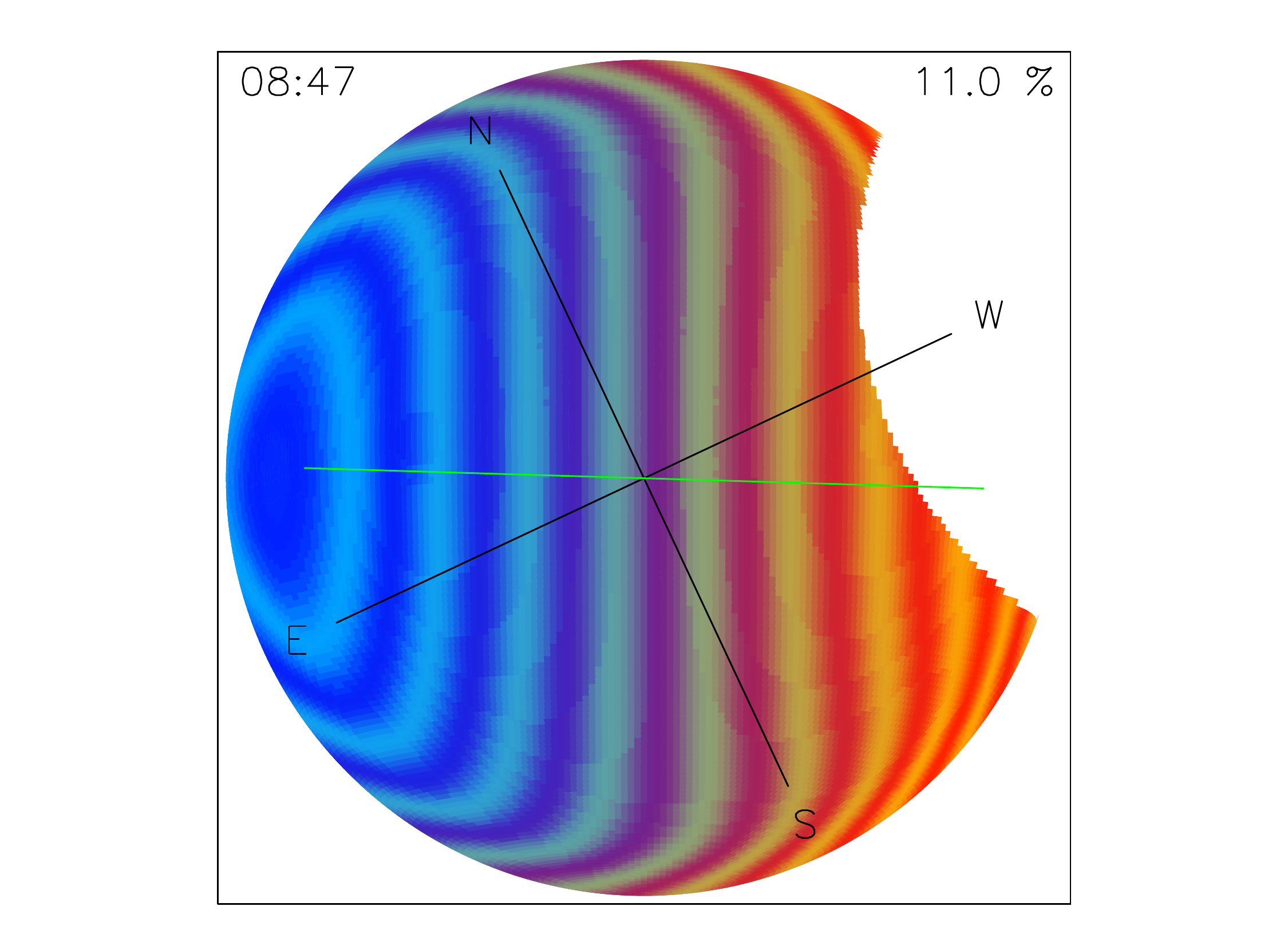}}
    \resizebox{.33\hsize}{!}{\includegraphics[bb = 107 12 540 445]{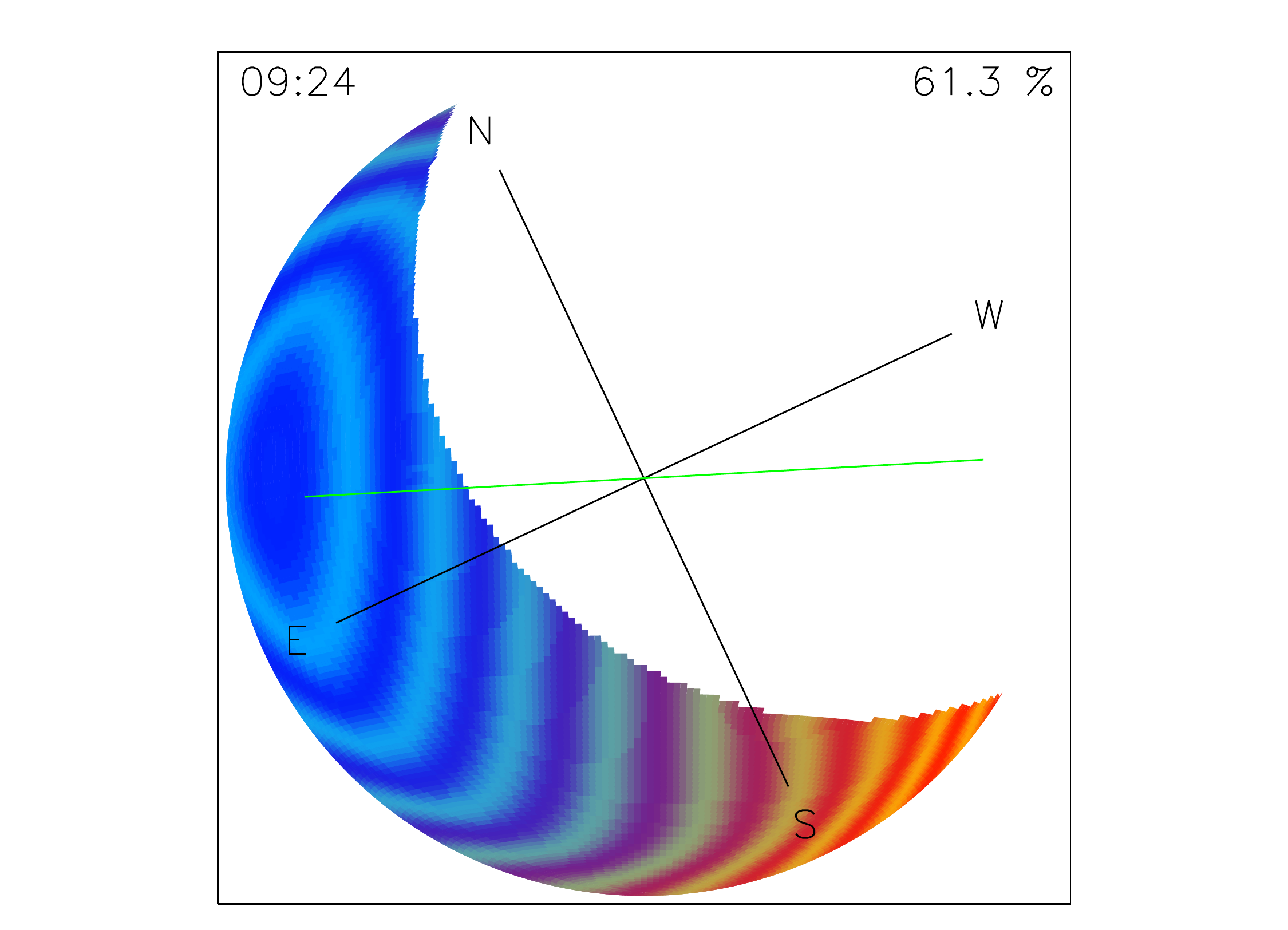}}
  }
  \mbox{
    \resizebox{.33\hsize}{!}{\includegraphics[bb = 107 12 540 445]{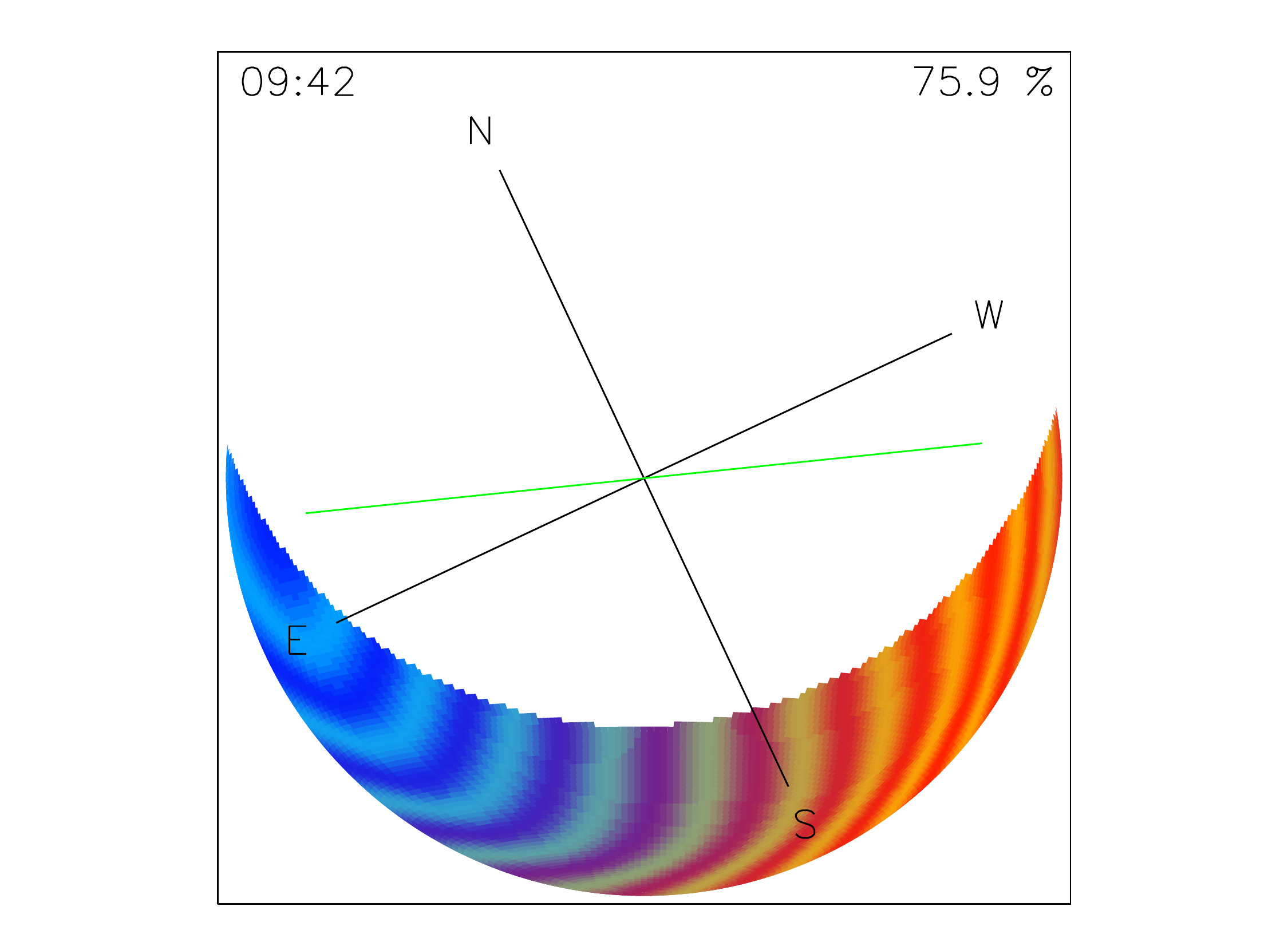}}
    \resizebox{.33\hsize}{!}{\includegraphics[bb = 107 12 540 445]{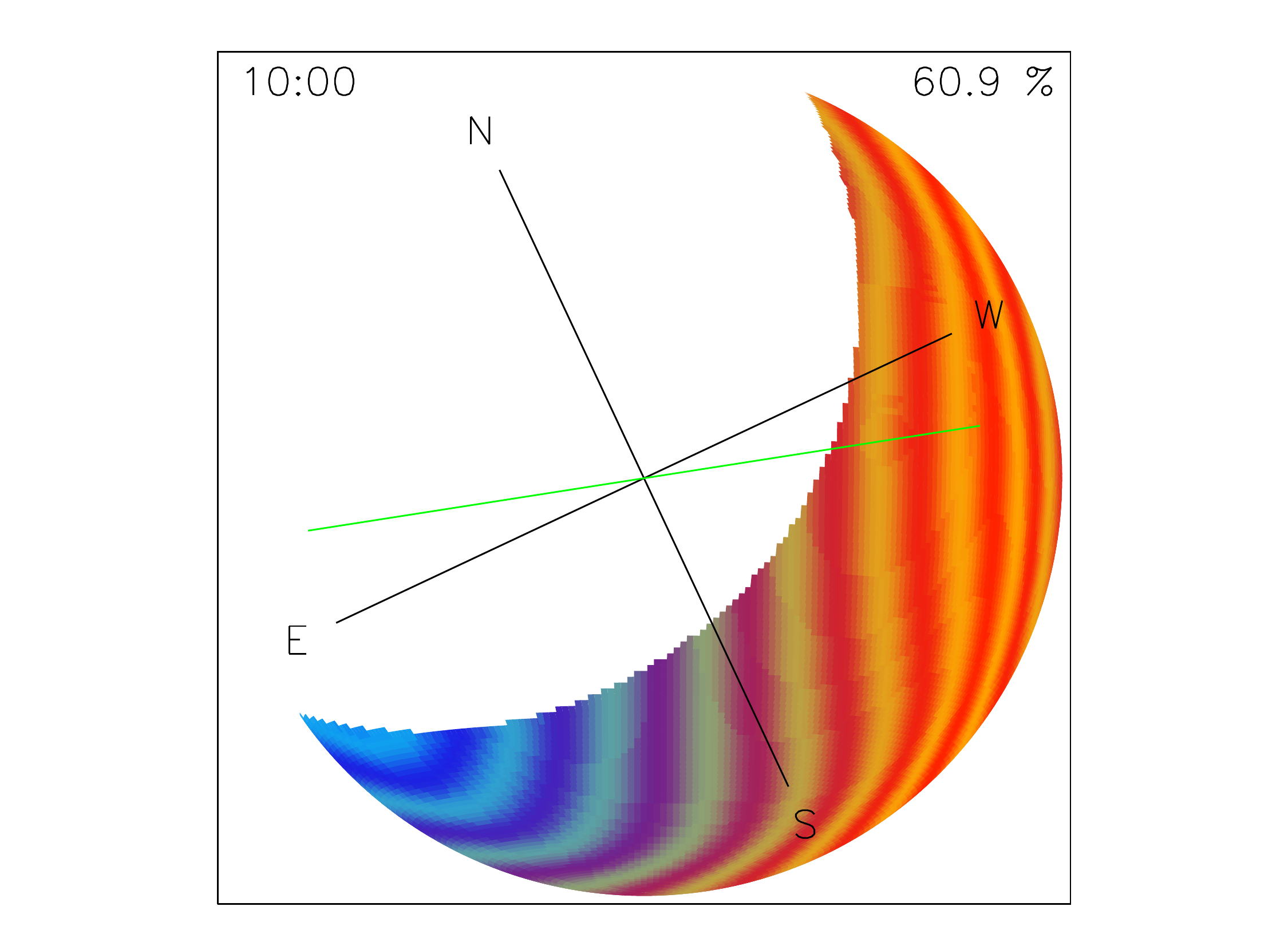}}
    \resizebox{.33\hsize}{!}{\includegraphics[bb = 107 12 540 445]{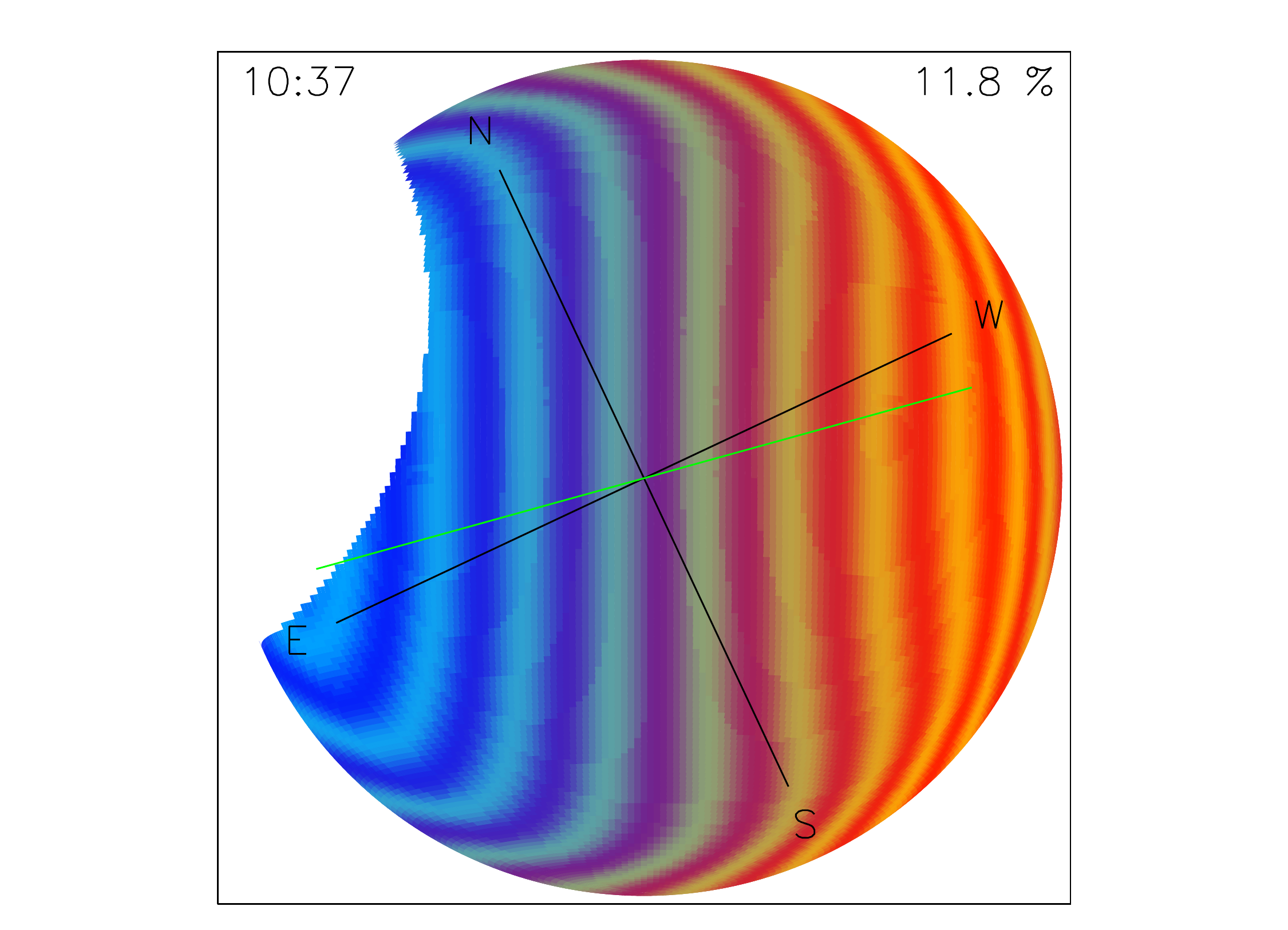}}
  }
  \caption{\label{fig:Eclipse}Eclipse configuration in synthetic
    Dopplergram. Colors show the velocity projected toward the observer;    the color coding is the same as in the middle panel of
    Fig.\,\ref{fig:SDO} with velocities between $-3$ and
    $+3$\,km\,s$^{-1}$ from blue to red with yellow stripes. Universal
    time and fraction of the surface occulted by the Moon are given in
    the upper left and upper right part of the images,
    respectively. Orientation on sky is shown as the black coordinate
    system, the green line shows the orientation of the horizon at any
    given time. White circles in the upper left panel show solar limb
    angles of $\mu =0.8, 0.6$ and $\mu = 0.3$.}
\end{figure*}

In Fig.\,\ref{fig:Eclipse} we show the configuration of the eclipse
on Mar\,20, 2015. The solar disk is shown at six different times with
projected radial velocity color-coded between $-3$ and
$+3$\,km\,s$^{-1}$ from blue to red with yellow stripes indicating
lines of equal RVs (color coding as in the center panel of
Fig.\,\ref{fig:SDO}). In the upper left corner of each panel, UT time is
shown. In the upper right corner, we provide the fraction of the solar disk
that is occulted by the Moon at each time of observation. From the upper left
to the lower right panel, the Moon crosses the solar disk, occulting
different parts of it. In the individual panel images, the solar
rotation axis is always oriented vertically (heliocentric). The
celestial coordinate system's N-S and E-W directions are indicated in
black. The projected orientation on the sky changes with time; the
orientation of the local horizon is shown as a green line. It changes
according to the path of the Sun across the sky. Doppler velocity and
convective blueshift are parameterized as explained above. At our
observatory in G\"ottingen, eclipse maximum was observed at 09:42\,UT
with an occulted area of almost 76\,\% of the solar disk.

The color-coding of radial velocity reveals an interesting feature:
the combination of latitudinal differential rotation and convective
blueshift leads to almost vertical lines of equal RV for most of the
central part of the solar disk. At low $\mu$-values, toward the limb
of the Sun, lines of equal RV show a strong gradient toward the left
in the plot, which is caused by the relative redshift at low $\mu$, that is,
velocities close to the limb become suddenly redshifted on the
projected solar disk (cf.\ Fig.\,\ref{fig:Blueshift}).

During eclipse, the Moon occulted the solar disk from the west, that is,
from the redshifted side of the velocity distribution. The RM curve
therefore first showed RVs distorted to the blue. This is different from
the canonical picture in stars that are orbited by planets whose
orbital motion is in the same direction as the stellar rotation. In
the Sun-Earth-Moon system, this is caused by the fact that all orbital
and angular velocities point in the same direction, but we observe
the Moon occulting the Sun while orbiting Earth.

\subsection{Solar surface observations}

An alternative, empirical description of the solar velocity fields and their magnetism can be derived directly from solar observations, for example,
from images taken by the Helioseismic and Magnetic Imager
\citep[HMI;][]{2012SoPh..275..229S} onboard the Solar Dynamics Observatory
\citep[SDO;][]{2012SoPh..275....3P}. We downloaded images of intensity,
Doppler velocities, and the magnetogram taken on March 20, 2015, at
09:36:00\,UT with HMI from
JSOC\footnote{\texttt{http://jsoc.stanford.edu}}. Exposure times were
720\,s. The images are shown in Fig.\,\ref{fig:SDO} at the typical resolution of
our RV calculations (see below).

In the intensity image, solar limb darkening and spots
are visible. The Dopplergram shows a similar structure as the
parameterized model, but also much more small-scale structure from
noise or granulation. The magnetogram reveals the magnetic areas of
opposite polarity near the equator. The magnetogram allows us to test
scenarios with convective blueshift suppressed at certain magnetic
field strengths. We discuss this in Sect.\,\ref{sect:Results}.

Dopplergrams were used before to infer information about the effect of
solar granulation and active regions on solar RV measurements
\citep[e.g.,][]{2010A&A...519A..66M, 2016MNRAS.457.3637H}. So far, it
has never been shown that the RV variations deduced from solar
Dopplergrams are quantitatively similar to disk-integrated RV
observations. We argue that HMI Dopplergrams are not a useful source
of information for accurate RVs because (i) Dopplergrams exhibit a
number of systematic effects across the solar disk that can introduce
errors of up to a few 100\,m\,s$^{-1}$ between different areas on the
disk; (ii) there is a significant systematic difference between the
observation of a stellar or solar disk-integrated RV and the
integration of Dopplergram RVs for individually resolved disk areas
(the canonical way to disk-integrate Dopplergram information is to use
intensity to weight individual surface areas. This approach neglects
the difference in line profiles across the solar disk, which
introduces a second mechanism of weighting between surface elements);
and (iii) the calculation of velocities across the solar disk relies
on a number of filters across one spectral line, but the line shape
varies across the disk, with the result that information from the limb
and disk center cannot directly be compared without introducing errors
much larger than several 10\,m\,s$^{-1}$. These effects are described
in \citet{2011SoPh..271...27F}, \citet{2012SoPh..275..229S},
\citet{2012SoPh..278..217C, 2012SoPh..275..285C}, and
\citet{2013ApJ...765...98W}.

A visual comparison between our synthetic Dopplergram (upper left
panel in Fig.\,\ref{fig:Eclipse}) and the observed HMI Dopplergram
(middle panel in Fig.\,\ref{fig:SDO}) reveals some striking
differences. First, the vertical yellow stripes only show very little
curvature in the synthetic images (Fig.\,\ref{fig:Eclipse}), but they are
significantly curved in the observed Dopplergram
(Fig.\,\ref{fig:SDO}). Most important is the difference very close to
the limb; at limb angles $<0.6$ the surface areas in the synthetic
Dopplergrams show a sudden increase in Doppler velocities to the red
(more positive velocities closer to the limb). This is visible in the
knee of the yellow stripes at around $\mu = 0.6$ and is caused by our
parameterization of convective blueshift seen in
Fig.\,\ref{fig:Blueshift}. Although the data appear more
fine-structured in the HMI Dopplergram, this trend clearly
is much weaker in the Dopplergram, meaning that the blueshift adopted in our
model is not entirely seen in HMI observations. We show below that the
solar RV curve confirms the relevance of this convective blueshift for
disk-integrated solar observations.

\subsection{Radial velocity calculation}

We calculated Sun-as-a-star radial velocities by integrating over the
visible surface disk. We divided the solar surface into segments of
equal area. For our calculations, we segmented the surface into 50,830
segments of equal area of which 25,417 were visible as long as the Sun
was not occulted by the Moon. The inclination of the Sun during our
observations was $i =
97.05\degr$.\footnote{\texttt{http://bass2000.obspm.fr/ephem.php}}

For each surface segment, we computed the velocity at the center of
the segment and weighted its contribution by the projected size and
intensity according to limb darkening and extinction. The segment's
projected velocity was calculated from the solar rotation law and
convective blueshift. At the angular size of the Sun of almost half a
degree, extinction and barycentric velocity significantly differ
across the solar disk. We computed the barycentric correction for each
individual segment with values of right ascension and declination
computed for the center of each segment. We verified that the
barycentric correction from JPL emphemeris for the center of the Sun
agreed with our results.

Extinction describes the effect of intensity attenuation caused by Earth's
atmosphere. Attenuation can be described with the formula
\begin{equation}
  I = I_{\rm 0} \cdot \, 10^{-(kx)/2.5},
\end{equation}
with $k$ the extinction coefficient and $x$ airmass; typical
extinction coefficients are between 0.1 and 0.6 for visual
wavelengths. We can estimate the effect of extinction on the solar RV
curve from Fig.\,\ref{fig:Eclipse}; extinction is stronger for regions
on the solar disk located at higher airmass, that is, below the green
lines. At maximum eclipse (09:42\,UT), the orientation of the horizon
was almost perpendicular to the rotation axis of the Sun, while the
line connecting Sun and Moon is almost parallel to the rotation
axis. This led to a degeneracy between the effects from extinction and
velocity fields during eclipse without consequences for our
calculations.

\begin{figure*}
  \centering
  \mbox{
    \resizebox{.33\hsize}{!}{\includegraphics[bb = 107 12 540 445]{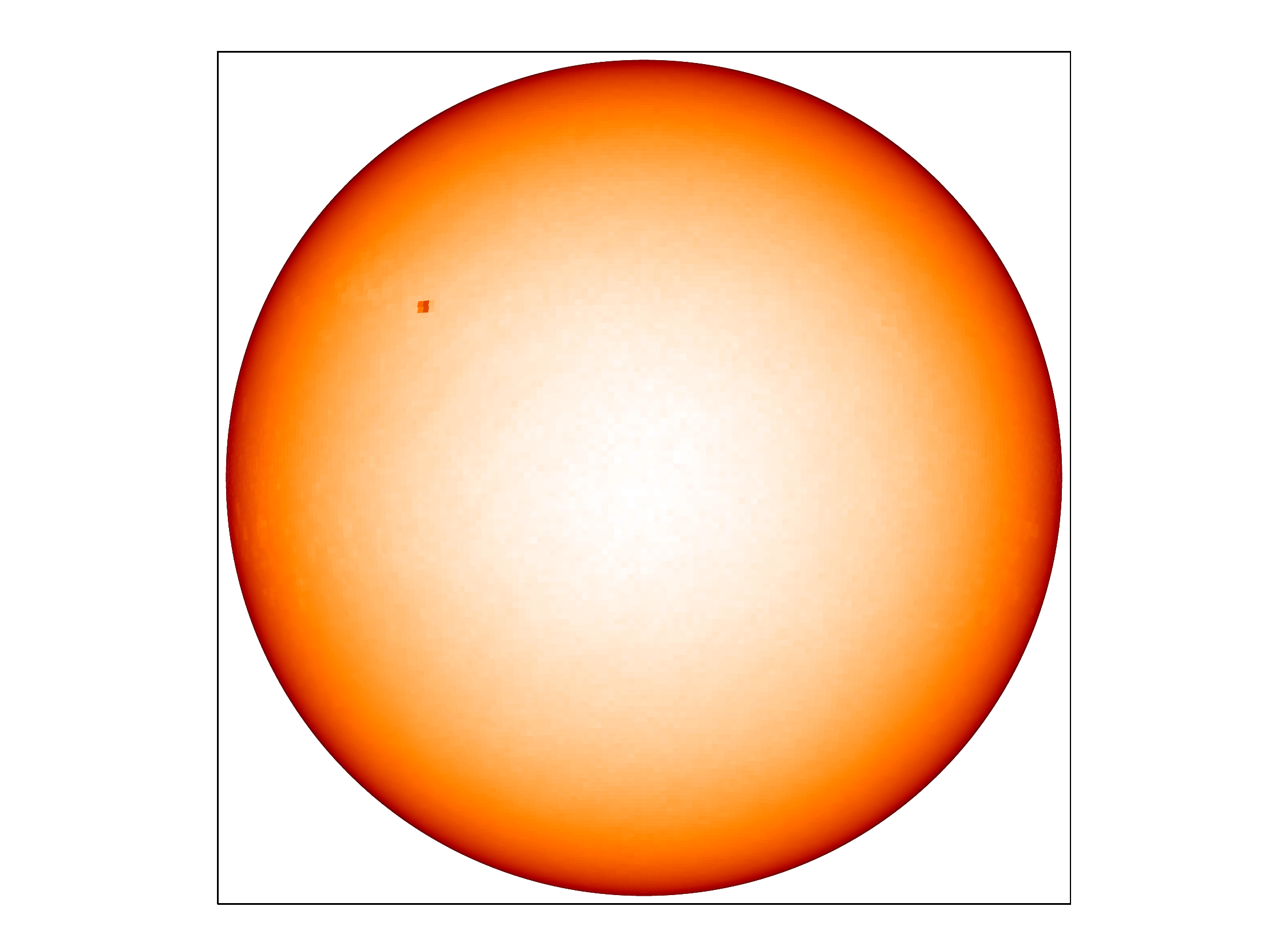}}
    \resizebox{.33\hsize}{!}{\includegraphics[bb = 107 12 540 445]{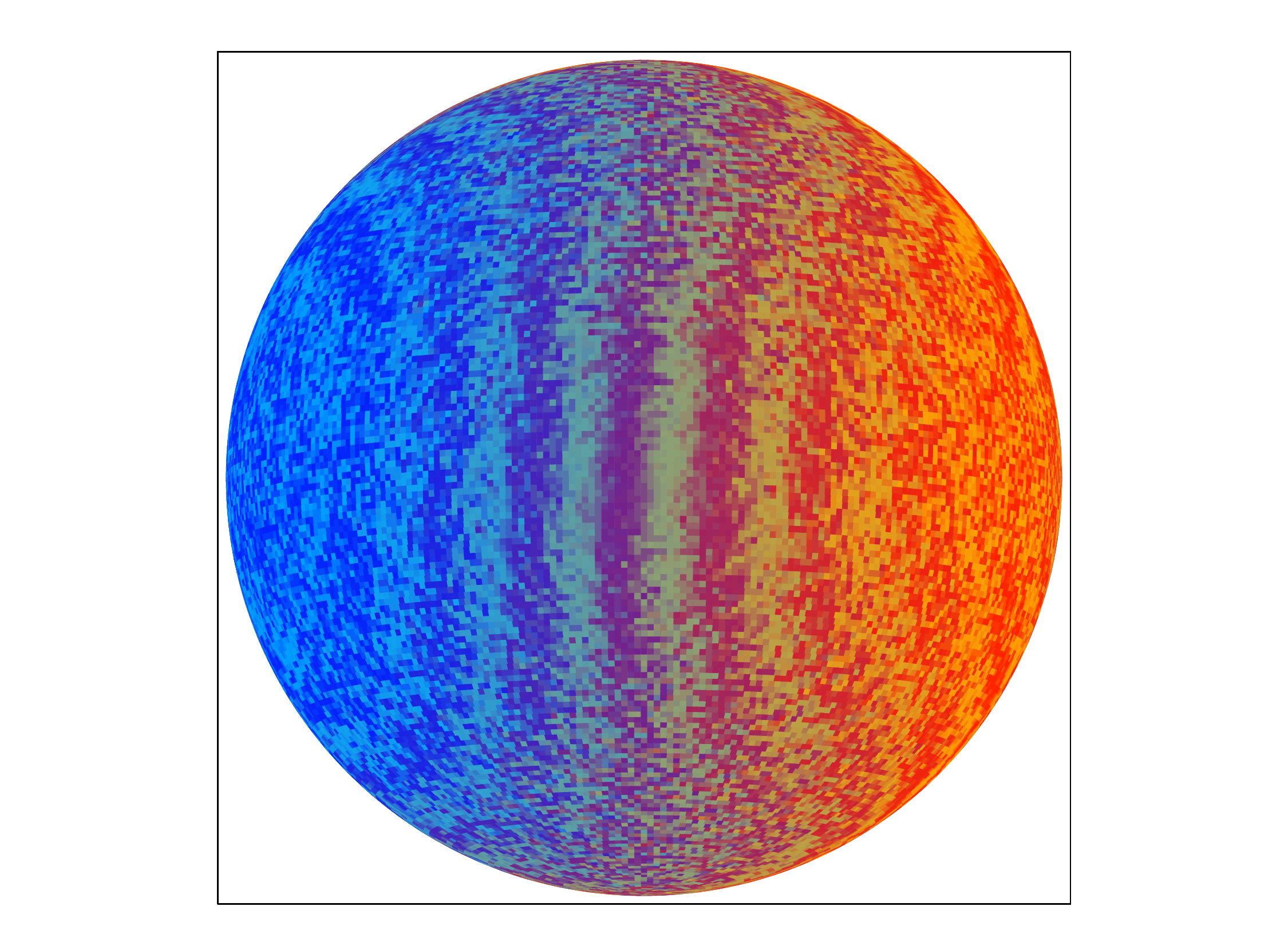}}
    \resizebox{.33\hsize}{!}{\includegraphics[bb = 107 12 540 445]{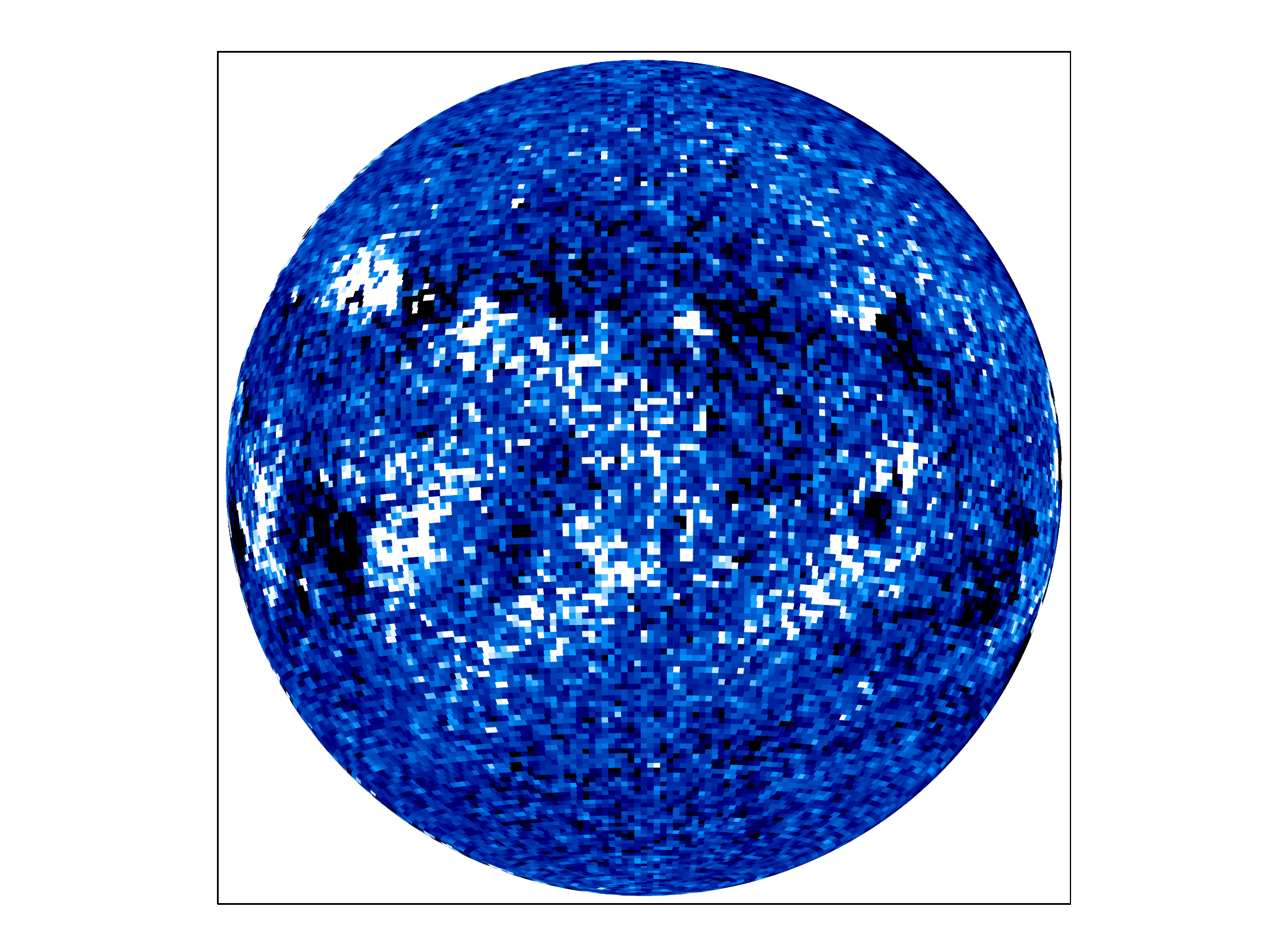}}
  }\vspace{9mm}
  \mbox{
    \resizebox{.33\hsize}{!}{\includegraphics[bb =  60 -20 648  45]{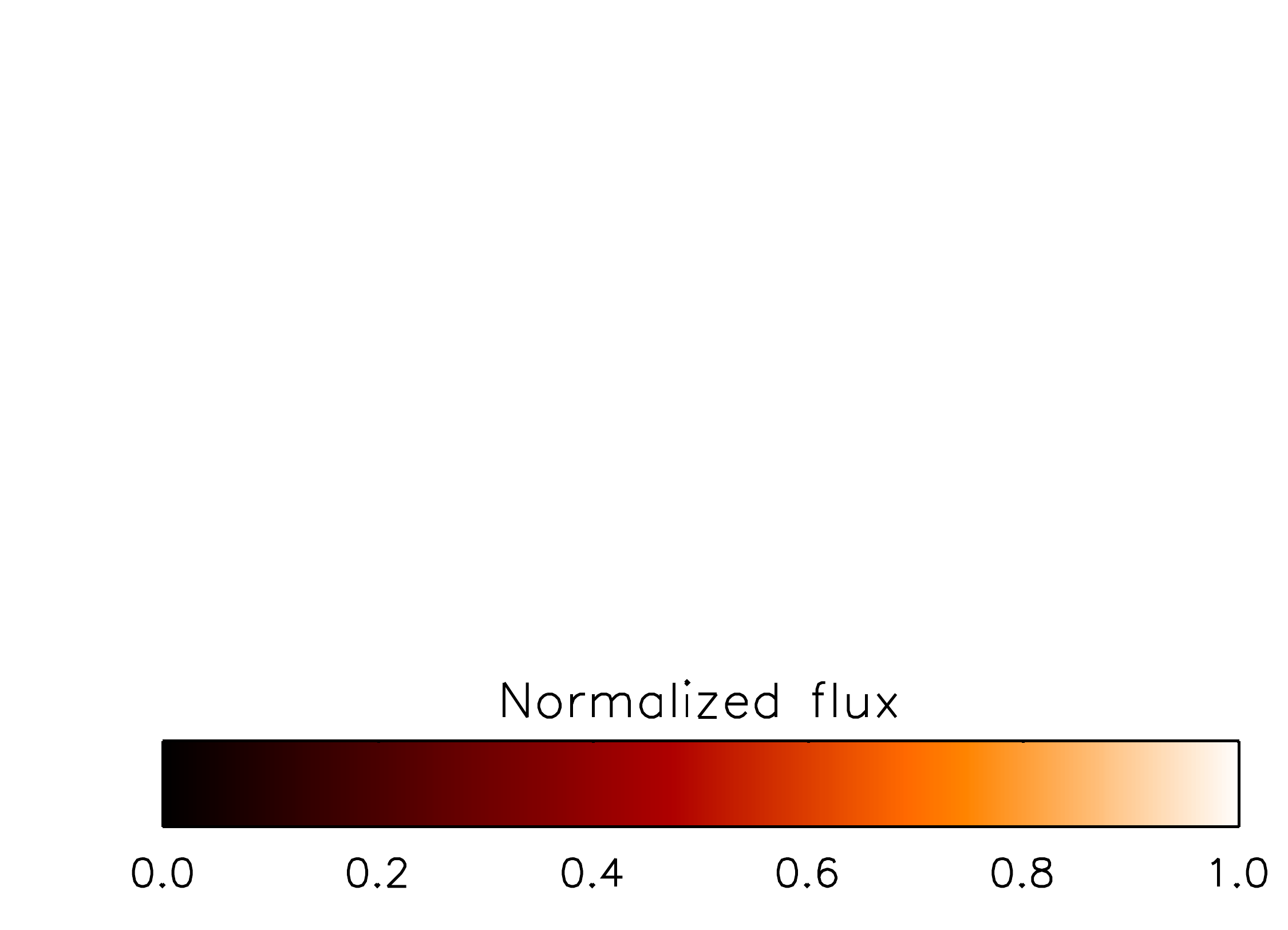}}
    \resizebox{.33\hsize}{!}{\includegraphics[bb =  60 -20 648  45]{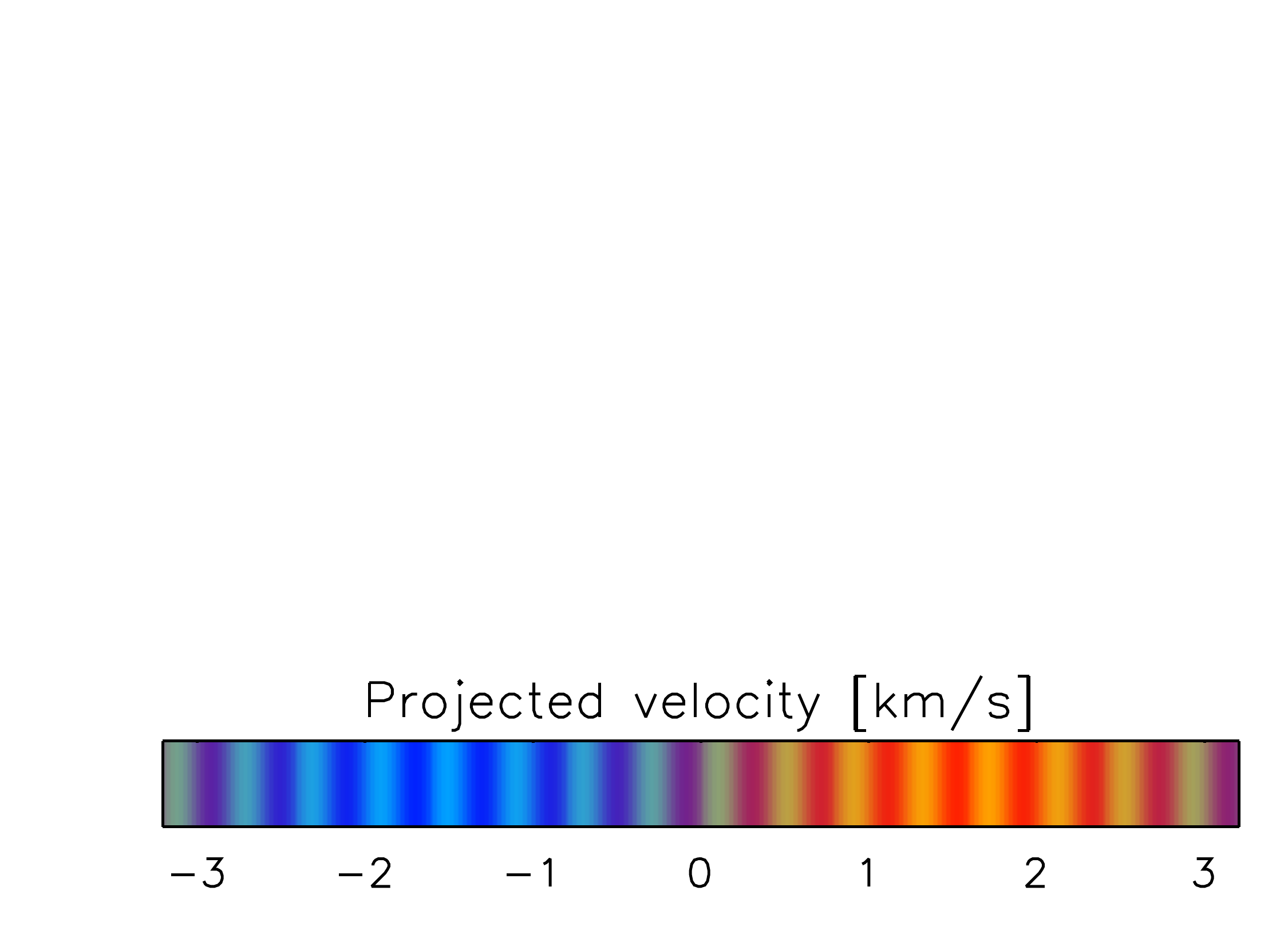}}
    \resizebox{.33\hsize}{!}{\includegraphics[bb =  60 -20 648  45]{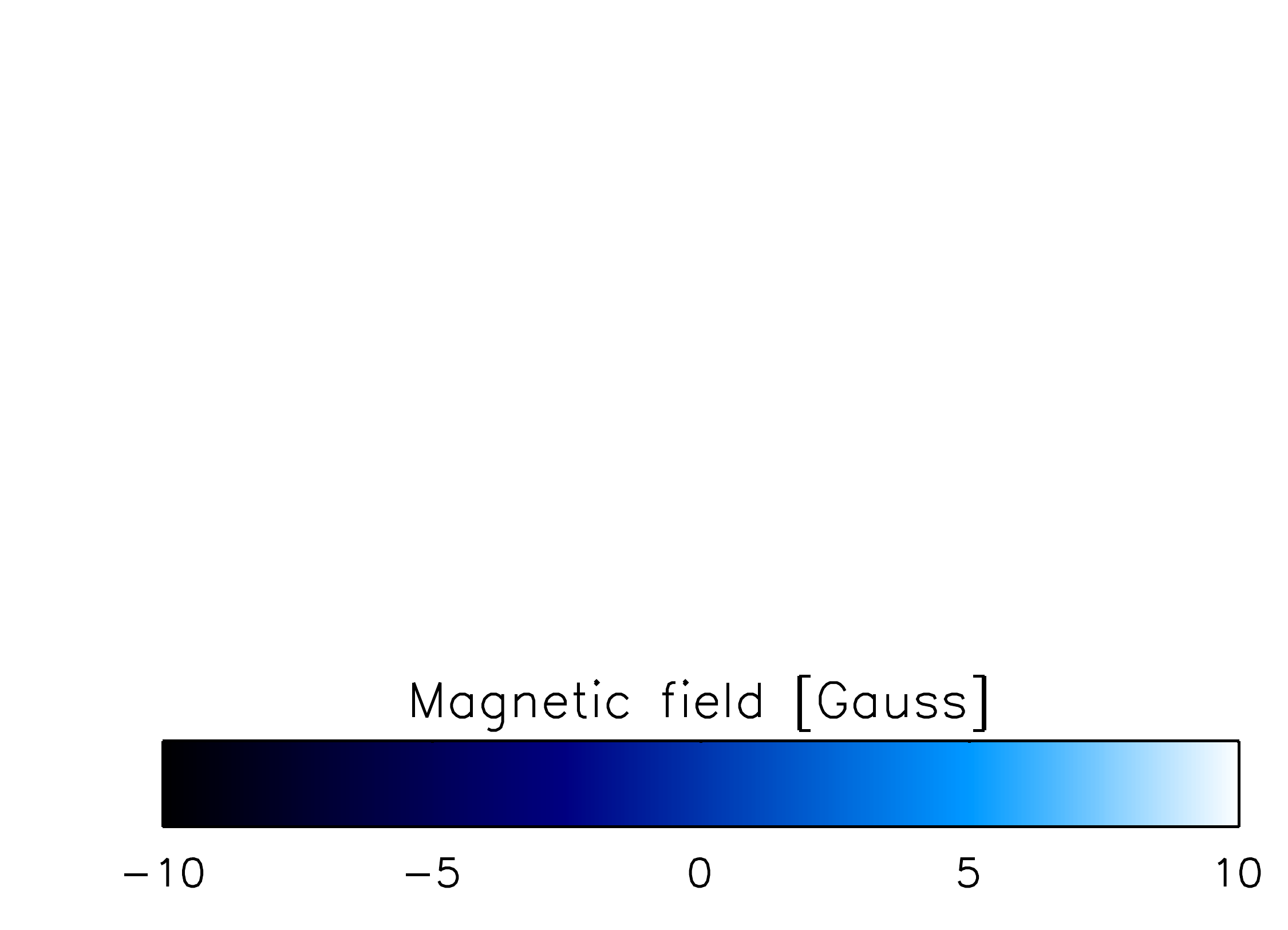}}
  }
  \caption{\label{fig:SDO}HMI images taken at 09:36\,UT, exposure time
    720\,s, shown at the resolution of our simulations. \emph{Left:}
    Intensity showing empirical limb darkening and a small sunspot at
    NE; \emph{Middle:} Dopplergram shown with the same color code as
    in Fig.\,\ref{fig:Eclipse}; \emph{Right:} Magnetogram showing
    positive (white) and negative (black) magnetic polarities.}
\end{figure*}

The value of the extinction coefficient is unknown but constrained by
the total intensity visible during the entire observation
(Fig.\,\ref{fig:Flux}). We tested the influence of extinction and
found that its value did not significantly influence the shape of the
RM curve (for realistic values of $k$). We used an extinction
coefficient of 0.4 for the beginning of the observations that linearly
diminishes in time to 0.1 at the end of observations. In the
wavelength range that we used to calculate the flux (5500 --
6000\,\AA), 0.1 is a reasonable value for good weather conditions as
during the end of our observations. At the beginning of our
observations, the Sun was just rising, and ice crystals
were probably still affecting extinction at very high airmass. Our model can
reproduce the observed flux in Fig.\,\ref{fig:Flux} very well,
differences like those between 08:00 and 08:30\,UT are probably caused
by variable extinction and/or thin clouds.

To compute the final RV, we can chose between different
strategies. The easiest (and by far the fastest) way is to calculate
the average of the radial velocities of all visible surface segments
weighted by their intensity, taking into account limb darkening and
extinction (as done in models~1, 2, and 4 below). A second strategy
is to compute the line profile for each segment, shift it according to
the segment radial velocity, add all the segment line profiles
weighted by intensity, and determine the RV of the final line profile
(model~3). This strategy can account for the problem that the RV
calculation may be affected by asymmetries in the final line profile
that can arise from the asymmetric velocity distribution of the
surface segments because of convective blueshift and the eclipse. It
can also take different line profiles for different solar
limb angles and intrinsically asymmetric lines into account. We tested both
strategies and present the results in the following.

\section{Results}
\label{sect:Results}

\subsection{Rossiter-McLaughlin curve}

We calculated the RM curve using four different models: model~1
averages velocities across the solar disk including no convective
blueshift; model~2 is the same, but includes convective blueshift.
Model~3 uses the full calculation of line profiles, and model~4
again is similar to models~1 and 2, but uses velocity fields taken
from satellite observations.

\subsubsection{Standard line model}

\begin{figure}
  \centering
  \resizebox{\hsize}{!}{\includegraphics[bb = 20 160 460 620]{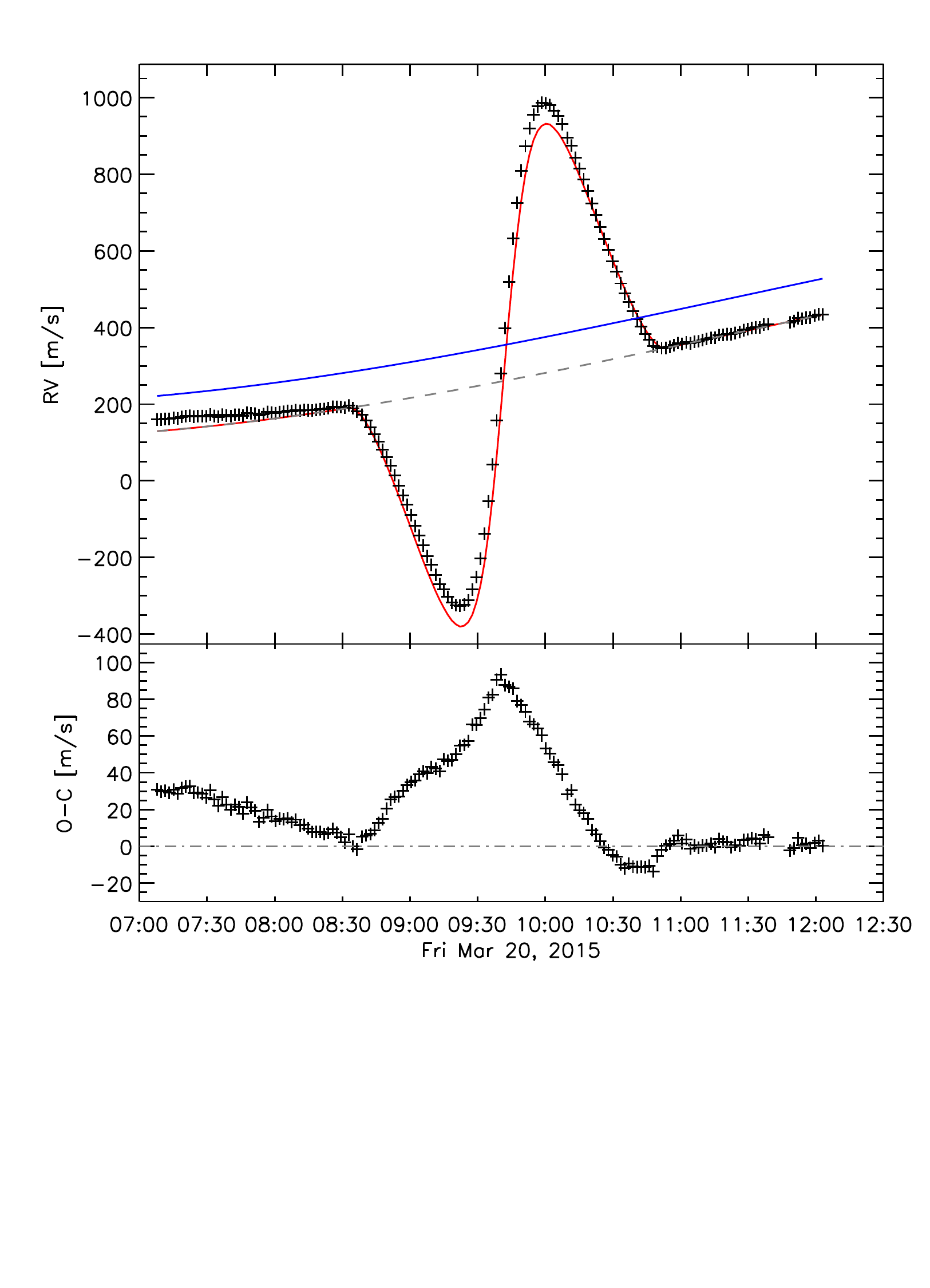}}
  \caption{\label{fig:RadCorr1}Observations of solar RVs during the
    eclipse together with the standard line model 1. \emph{Top
      panel:} black plus symbols show our solar RV observations. The
    red line shows results from our model corrected for an offset so
    that they match observations after the end of the eclipse (see
    text). The gray dashed line shows the eclipse-free model (only
    barycentric motion). The blue line shows barycentric motion from
    JPL's ephemeris. \emph{Bottom panel:} Residuals between
    observations and model (black symbols minus red curve in top
    panel).}
\end{figure}

Our first calculation of the RV curve, model 1, follows the simplest approach; we parameterized limb darkening and solar
differential rotation as introduced in
Section\,\ref{sect:parameterization}, and we calculated RVs from the
weighted velocities of all visible segments, that is, we did not compute
any line profiles. In this first attempt, convective blueshift was not
taken into account. We show the results of this calculation and a
comparison to our observations and to solar barycentric velocities
from JPL's ephemeris in Fig.\,\ref{fig:RadCorr1}: in the top panel, we
show our RV observations (black symbols) together with the results
from the first model (red line). The dashed gray line shows our model
results for the unobscured solar disk (no eclipse). The blue line
shows solar velocities as taken from JPL's ephemeris. For our plots,
we decided to show the JPL emphemeris and our solar RV observations with
no additional RV offset. However, for better comparison between our
eclipse model (red line) and solar RV observations, we applied an
offset to the eclipse model. In model 1, without convective
blueshift, our eclipse by design reproduces the JPL ephemeris. The
offset between the two is only 30\,cm\,s$^{-1}$. The offset between
JPL ephemeris and solar RV observations is 94\,m\,s$^{-1}$, hence the
solid red and dashed gray lines in Fig.\,\ref{fig:RadCorr1} are offset
by that amount. This offset is partly caused by inaccuracies of our
optical system \citep[see][]{2016arXiv160300470L}, but we expect a
systematic error of only a few ten m\,s$^{-1}$. The bulk of the
difference between ephemeris and solar observations (94\,m\,s$^{-1}$)
is caused by the real difference between the two mechanisms that
affect integrated solar RV observations: net convective blueshift and
gravitational redshift. Our observations indicate that the sum of both
effects is on the order of 90--100\,m\,s$^{-1}$ , but the uncertainties
are large (see above). This value could be determined with higher
accuracy if the systematic uncertainties of our optical setup were
better understood. Such a measurement would allow the direct
determination of the net convective blueshift because the
gravitational redshift is known to relatively high accuracy
\citep{2003A&A...401.1185L}.

The comparison between observations, JPL ephemeris, and our model
calculations shows that the main ingredients of the eclipse RV
observations are captured by our model. The barycentric motion of the
Sun varies by roughly 300\,m\,s$^{-1}$ during our observations. The RM
effect produces an asymmetric curve with amplitudes between
$-500$\,m\,s$^{-1}$ and $+800$\,m\,s$^{-1}$ in addition to barycentric
motion. In this first model without convective blueshift, the
calculated RM curve appears systematically offset toward the
blue. The bottom panel of Fig.\,\ref{fig:RadCorr1} shows the residuals
between RV observations and the model results. The difference during
transit is clearly visible, reaching almost 100\,m\,s$^{-1}$ during
eclipse center. We ascribe the main part of this difference to
convective blueshift.

During the first 1.5\,hours before the eclipse, our RVs show a
systematic trend that is visible as residual starting at
30\,m\,s$^{-1}$ at 07:00\,UT and approaching zero difference after
08:30\,UT, just when the eclipse begins. Such an effect is not visible
after the eclipse, residuals after 11:00\,UT scatter closely around
zero. We ascribe this effect to the systematic problem of our optical
setup. We showed in \citet{2016arXiv160300470L} that RV trends of this
amplitude were found in other observations, probably caused by a
systematic problem with the way we couple sunlight into the fiber. The
linear slope of the residual before 08:30\,UT is very similar to what
we see in other observations. Unfortunately, we cannot exclude that
similar effects occurred during eclipse or after, but we only expect
effects that would be detected as linear trends or sudden jumps in the
residuals. We do not see clear evidence of such effects. Thus, we
assume for the following that the trend we see before 08:30\,UT is
caused by our optical setup and that no such effect occurs later
during observation. It is suspicious that the trend stops exactly at
the beginning of the eclipse, but the residuals give no reason to
believe that there is a significant effect at any time later during
our observations. In the comparisons to our other models,
we correct for this linear slope from now on by division through a straight fit to
the data before 08:30\,UT.

After correcting for this trend, we were able to measure the scatter of our
solar RV measurements outside eclipse. We found that before and after
eclipse (81 observations) our RV values scatter around the barycentric
trend with an rms of 2.2\,m\,s$^{-1}$. The cadence of these
observations is 97\,s, which means that part of this scatter is
probably caused by solar oscillations. After averaging our
measurements into $\sim5$\,min bins (averaging three exposures, 27 bins),
we found that our RV measurements scatter with an rms of
1.6\,m\,s$^{-1}$. Further binning reduced the scatter even more:
binning five exposures into 16 data points yielded an rms of
1.25\,m\,s$^{-1}$, and ten exposures with 8 data points yielded
0.7\,m\,s$^{-1}$. The scaling of the rms scatter with bin size follows
expectation from purely statistical scatter. We conclude that our
data points outside eclipse are not systematically affected by trends
other than the linear trend discussed above. The statistical nature of
the RV scatter could be caused by solar short-term variability or by
the statistical noise in our RV measurements. Our estimate from photon
noise is on the order of only 10\,cm\,s$^{-1}$. We therefore suspect
that the source of the RV ``noise'' in our observations of the Sun as
a star is in fact solar granulation.

\subsubsection{Standard line model with convective blueshift}

\begin{figure}
  \centering
  \resizebox{\hsize}{!}{\includegraphics[bb = 20 160 460 620]{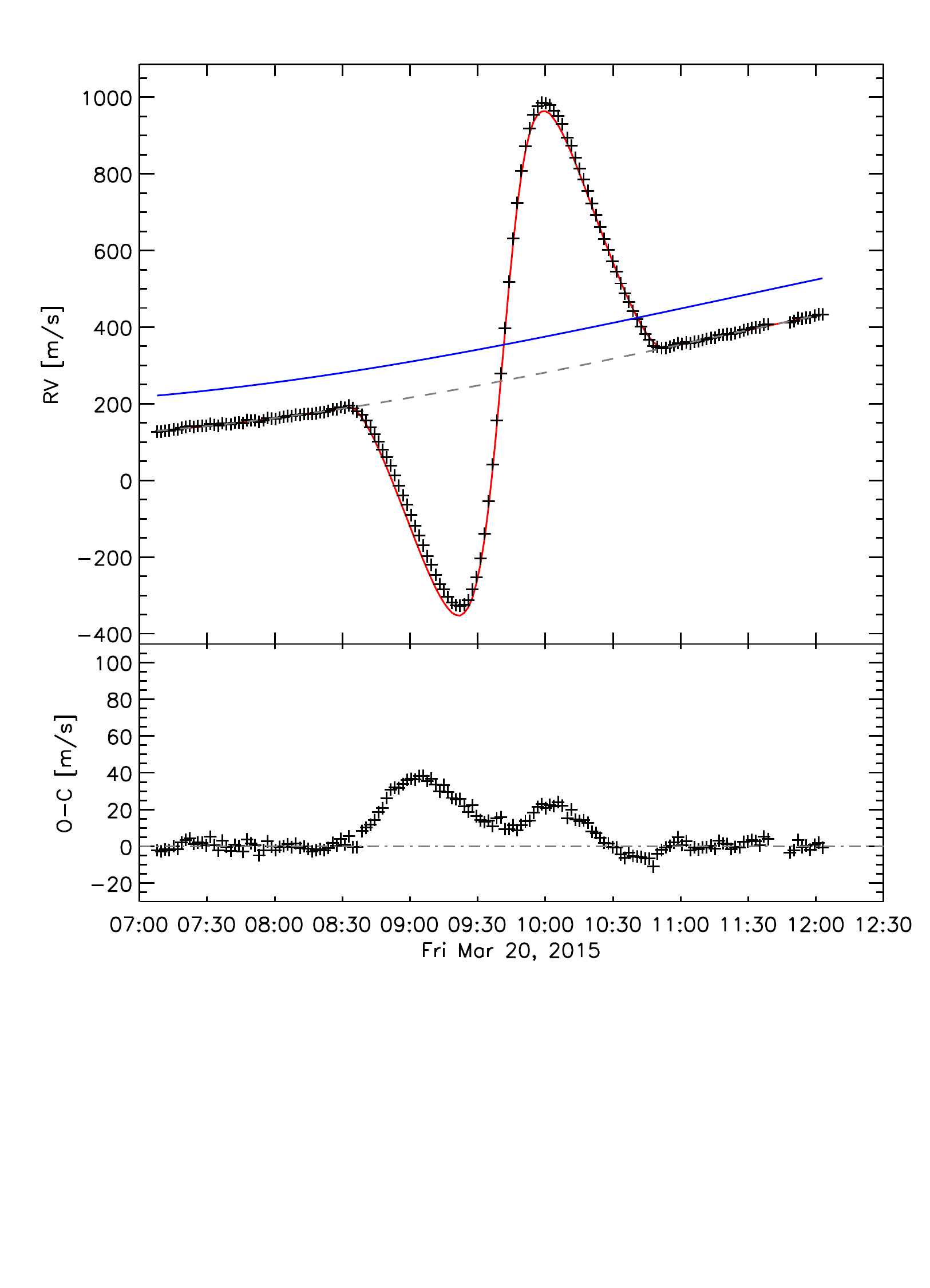}}
  \caption{\label{fig:RadCorr2}Observations of solar RVs during the
    eclipse together with results from model 1 including convective
    blueshift. Lines and symbols are the same as in
    Fig.\,\ref{fig:RadCorr1}. Here, the model velocities (red line)
    match the solar observations without any offset.}
\end{figure}

In model 2 we used the same strategy to calculate solar RVs as in
model 1, but added the parameterization of convective blueshift as
shown in Fig.\,\ref{fig:Blueshift}. The results are shown in
Fig.\,\ref{fig:RadCorr2}. Lines and symbols are the same as in
Fig.\,\ref{fig:RadCorr1}, the model RVs (red line) are not offset, but
their absolute velocities match the solar RV observations because
the offset is one parameter for the blueshift velocities. As explained
above, we do not claim that our parameterization provides absolute
values for the blueshift because there may be systematic offsets in
our observations. More important than the absolute RV calibration is
that the calculated RVs provide a significantly better match to the
observations than the model without blueshift; phenomenologically, the
modeled RM curve is shifted to redder velocities. The residuals
between calculated and observed RVs during eclipse are still fairly
high, up to 40\,m\,s$^{-1}$ early during eclipse, but they are lower
later.

We noted in Sect.\,\ref{sect:Instrument} that the time of our
observations is unknown within several 10\,s because of an offset in
our instrument clock. This means we have the freedom to shift the
observations in time by several 10\,s. Boundary conditions for the
time correction are that the calculated values must match the flux
variations as shown in Fig.\,\ref{fig:Flux} \emph{and} the RVs. There
is some degeneracy between the time and the amount of blueshift at
least during the center of eclipse. Better information comes from the
peaks in the RM curve and the time of ingress and egress. We have
extensively tested other parameterizations of blueshift and
combinations of blueshift and time-offsets; our parameterization of
blueshift together with an offset of 50\,s applied to the time of our
observations provides the RV curve as shown. We were not able to
produce a significantly better match to the data using any other set
of parameters. We conclude that the introduction of convective
blueshift is key to the understanding of the solar eclipse RV
curve. Nevertheless, residuals are still significant, and we see no way
to produce better RV calculations with this type of model.

\subsubsection{Line profile model}

\begin{figure}
  \centering
  \resizebox{\hsize}{!}{\includegraphics[bb = 20 160 460 620]{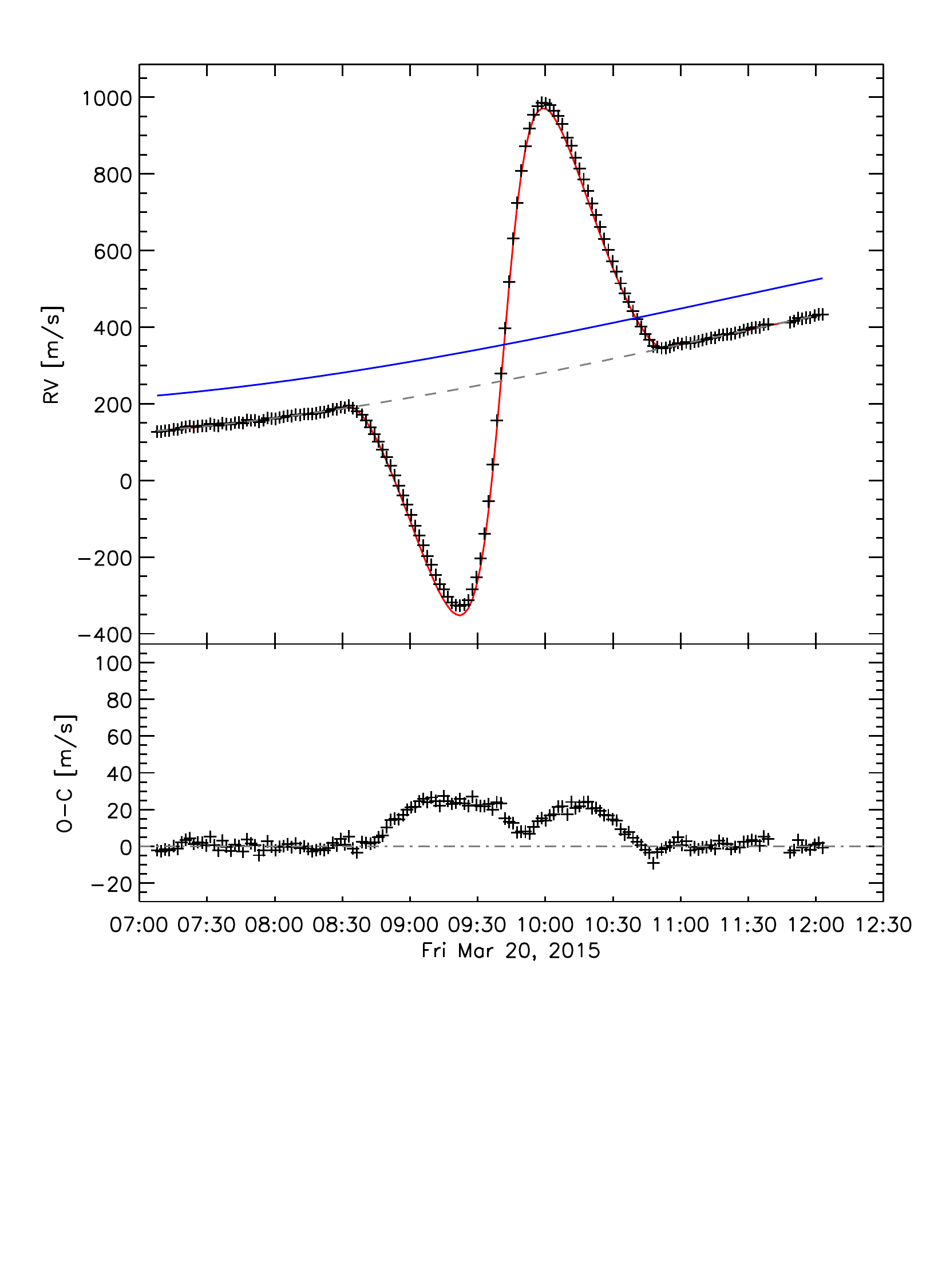}}
  \caption{\label{fig:RadCorr3}Observations of solar RVs during the
    eclipse together with results from model 3 involving the
    calculation of line profiles and determination of RVs from the
    profiles. }
\end{figure}

In our third model we calculated specific line profiles for every
segment on the visible solar surface. For that, we interpolated ten
synthetic line profiles of the Fe~I line at 6173\,\AA\ from
\citet{2013A&A...558A..49B} calculated at values of $\mu = 0.1, 0.2,
\dots 1.0$, and assigned a line profile to every surface segment (see
also Sect. \,\ref{sect:LineProfile} and
Fig.\,\ref{fig:SpecComparison2}). We shifted every line profile
according to the rotational Doppler velocity of each segment and
calculated a disk-integrated line profile with respect to limb
darkening and extinction. From this profile, we calculated the RV as
done for the solar observations. The potential advantages of this
approach are that line asymmetries and convective blueshift as
functions of limb angle are correctly included and that any
systematic effects from the way how RVs are calculated are captured.

The results from model 3 are shown in Fig.\,\ref{fig:RadCorr3} in
the same fashion as before. The mean velocities from our calculations
are offset by $-130$\,m\,s$^{-1}$ with respect to the solar RV
observations. This is not critical because the MHD simulations were not
designed to absolutely match observed solar wavelengths. The residuals
between the calculated RV curve and solar observations are somewhat
lower than the residuals calculated from the simpler model 2 (a
maximum of 30\,m\,s$^{-1}$ instead of 40\,m\,s$^{-1}$), but the
qualitative discrepancy during eclipse still exists. The relative
mismatch is always smaller than 10\,\%, which is a reasonably good
match. Nevertheless, the calculation of more realistic line profiles
and the calculation of RVs from profiles instead of simple weighted
velocities did not significantly improve the final RV model. This may
not be surprising because our parameterization of convective blueshift
in model 2 was derived from the line profiles we used in model 3,
and therefore most of the information was already included. On the
other hand, it implies that the way the RVs are calculated is not the
reason for the discrepancy and that a probably more fundamental
problem exists in our description of the solar surface spectra across
the solar disk. As a main candidate, we suggest the shape of line
profiles as a function of solar limb angle $\mu$.

\subsubsection{Solar velocity fields from HMI observations}

\begin{figure}
  \centering
  \resizebox{\hsize}{!}{\includegraphics[bb = 20 160 460 620]{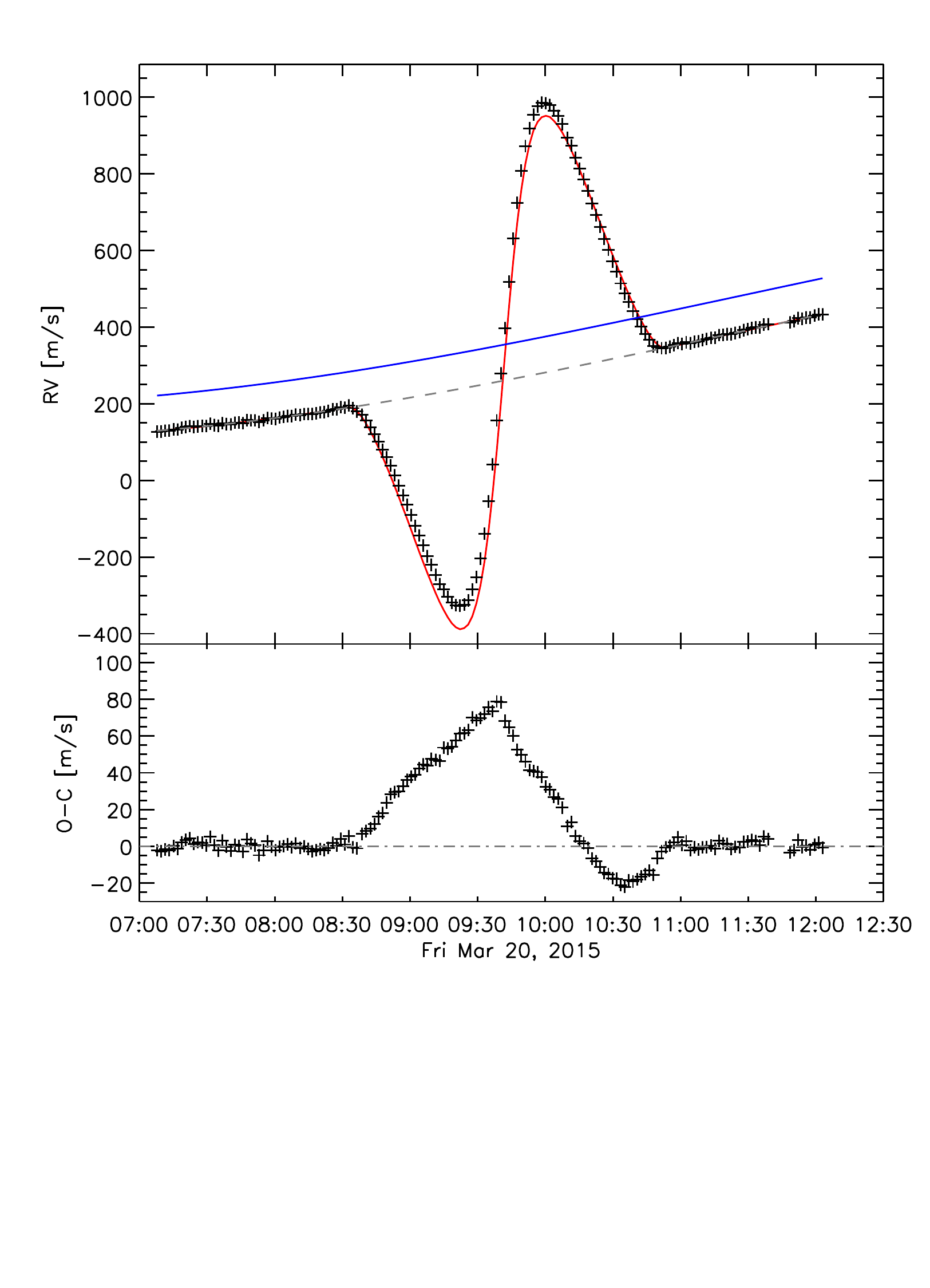}}
  \caption{\label{fig:RadCorrHMI}Observations of solar RVs during the
    eclipse together with results from model 4 adopting limb
    darkening and solar surface velocities from HMI
    observations. Lines and symbols are the same as in
    Fig.\,\ref{fig:RadCorr1}.}
\end{figure}

In our fourth model, we adopted limb darkening and solar velocity fields from
HMI observations (Fig.\,\ref{fig:SDO}), RVs were calculated as in our first
two models. With this model, we wished to test the information contained in HMI
Dopplergrams. We inspected the limb-darkening law seen in the HMI images and
confirm that it is very similar to the parameterization we used earlier. It
made no significant difference whether we used information on limb darkening
from the HMI images or from the parameterization. For consistency, we used the
image information in model 4. The resulting RV curve is shown in
Fig.\,\ref{fig:RadCorrHMI}, it is very similar to our first model, in which
limb darkening and rotational velocities were parameterized, but \emph{no}
convective blueshift was applied. This could either mean that our
parameterization of convective blueshift is incorrect, or that HMI Dopplergrams do
not capture the information relevant for full-disk surface integration. Our convection model appears to be the better choice because (a) we obtain a
much better match to the observed solar RM curve, and (b) the results from
full line-profile calculations (model 3) are very similar to the RVs,
including our convection model.

Our test in model 4 revealed a relevant result: it shows that when the
HMI Dopplergram is used to calculate an intensity-weighted
surface-integration to construct disk-integrated solar RVs, the result
is similar to including only solar surface differential rotation, but
it lacks most of the information about convective blueshift. This may
be partly due to the uncertainty of Dopplergram velocities, in
particular when Doppler velocities are compared across the full solar
disk \citep{2012SoPh..278..217C, 2012SoPh..275..285C,
  2013ApJ...765...98W}. A more important explanation probably is that
the way the Dopplergrams are constructed differs from the definition
of RVs as used in our model calculations \citep{2012SoPh..275..229S};
Dopplergram velocities are calculated from several filters around the
Fe~I line at 6713\,\AA, which implies that they reflect a certain velocity
field according to the effective formation height, which may not
coincide with the height we observe in our data
\citep{2011SoPh..271...27F}. To test the reliability of our
spectroscopic RV data and its relation to the 6173\,\AA\ line, we
computed RVs from our spectra using only this line. The results were
of course more noisy, but we found excellent agreement between results
from our broad-band spectra and those using only this single line. We
conclude that the choice of the line does not cause any bias in the
comparison between our RV determination and the HMI Dopplergrams, and
that the Dopplergram velocity fields, if used in this way to construct
RVs, capture the differential rotation of the Sun quite well, but not
the pattern of convective blueshift.

Another test we performed involves information about the solar magnetic
field. It is often assumed that magnetic fields stronger than a particular
value suppress convective blueshift and that relevant RV variations in solar
observations can be explained by blueshift suppressed in sunspots. For
long-term observations, this works through corotation and evolution of
magnetically active regions, but long-term evolution of RVs and solar active
regions cannot be assessed with our observations. Our attempts to improve the
RV model by applying lower blueshifts in regions of high magnetic fields
were all fruitless; we found no improvement to our RV curve when we modified the
blueshift pattern in regions of high magnetic fields. In particular, varying
the blueshift in the region of the visible spot did not significantly
influence the RV curve.

\subsection{Line profiles}
\label{sect:LineProfile}

Our observations equipped us with a way to compare our synthetic line
profiles to empirical data not only for spectra from the full
disk-integrated solar surface, in which much of the limb angle
information is lost, but also for spectra integrated over only parts
of the disk while other parts were eclipsed by the Moon.

\begin{figure}
  \centering
  \resizebox{\hsize}{!}{\includegraphics[bb = 30 10 643 445, clip=]{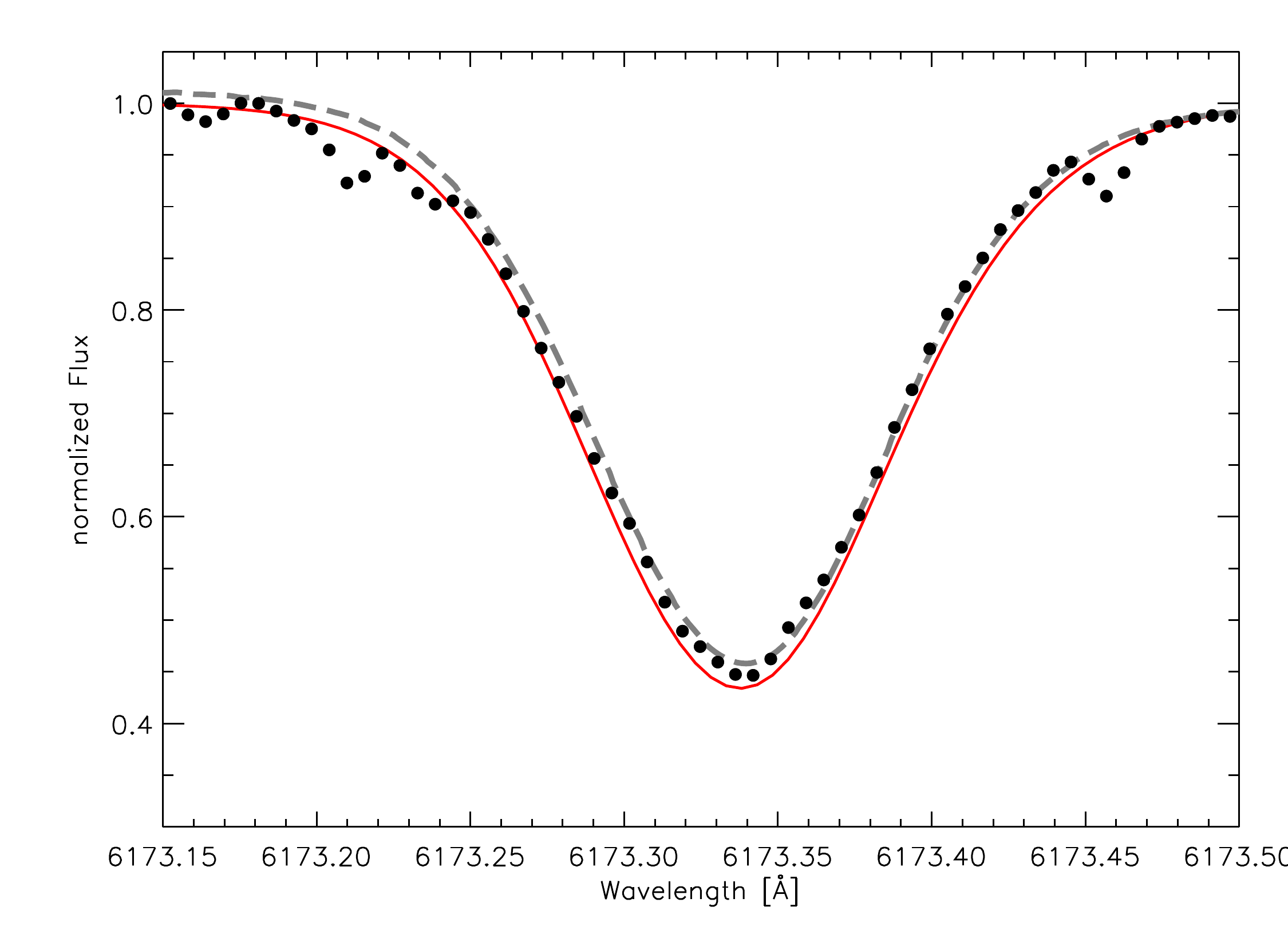}}
  \caption{\label{fig:Spectrum}Comparison of unperturbed line
    profiles; black circles: observed solar spectrum outside transit;
    gray dashed line: solar atlas; red solid line: modeled line
    profile for the entire solar disk. Telluric lines appear at
    $\lambda = 6173.2$--$6173.25$\,\AA\ and 6173.45\,\AA\ in the
    observations but are averaged out in the solar atlas.}
\end{figure}

A comparison between the fully disk-integrated spectrum for the
observed Sun and the line profile model for the 6173\,\AA\ line (from
model 3) is shown in Fig.\,\ref{fig:Spectrum}. In this figure we
show a calculated profile together with a spectrum taken during our
observations before transit with black dots, and the line from the
solar atlas of \citet{2016A&A...587A..65R}. The latter was constructed
from many individual exposures taken at very different barycentric
velocities. The quality of this spectrum is much higher, and it is
virtually free of telluric lines. The two observed solar spectra
compare very well within the uncertainties. As a red solid line we show the
result of the disk-integration of our 3D-MHD synthetic line
profiles. The differences between model and observations are small but
significant, they are on the order of 0.02--0.03 in normalized flux
units. There is an obvious difference between the overall line shape
of the observed and modeled profiles, which probably stems from larger
differences between real and modeled line profiles at certain
positions on the solar disk.

\begin{figure}
  \centering
  \resizebox{\hsize}{!}{\includegraphics[bb = 25 5 463 630, clip=]{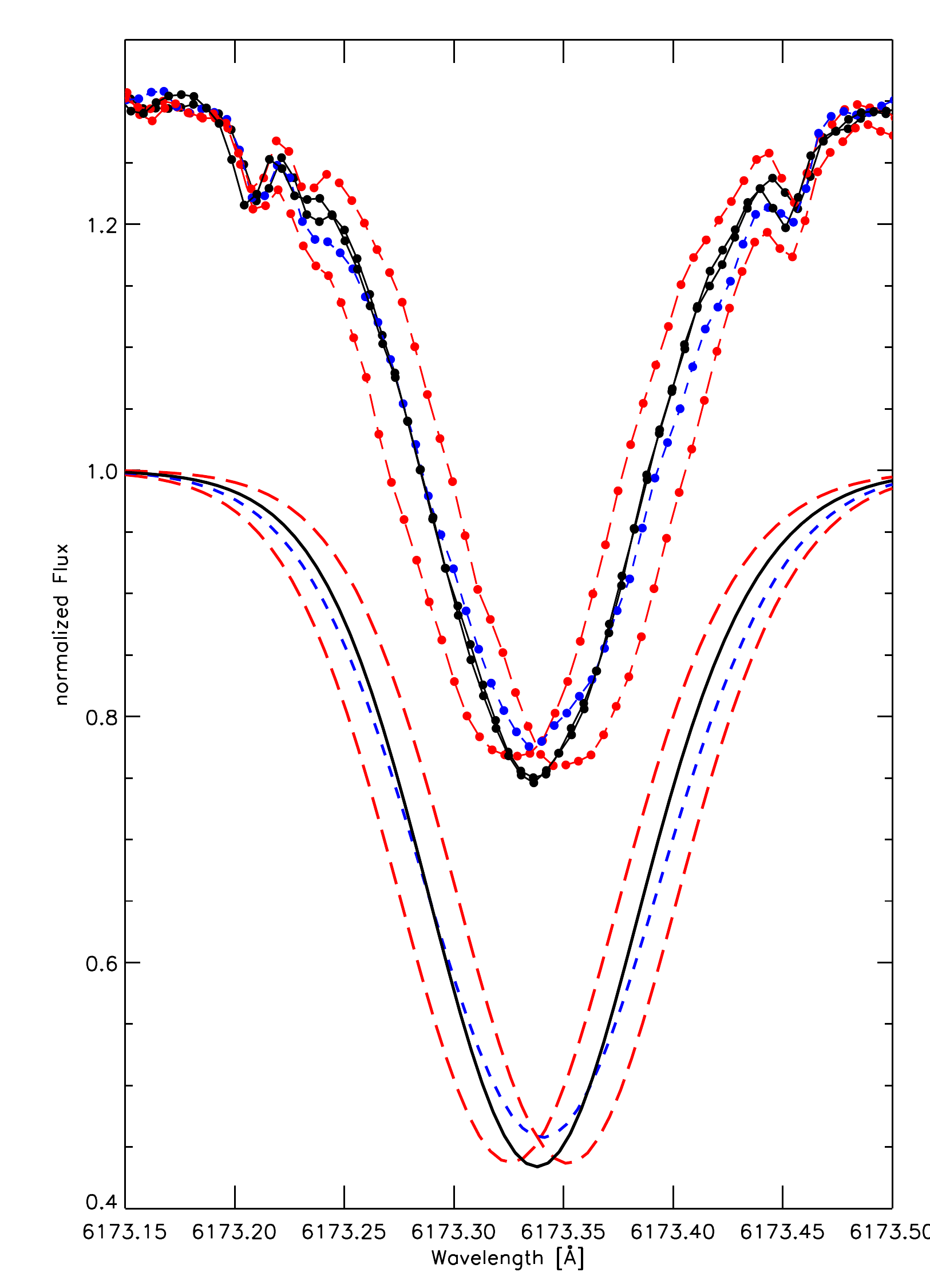}}
  \caption{\label{fig:SpecComparison1}Comparison between observed (top
    panel) and modeled (bottom panel) line profiles. In each panel,
    spectra at four different phases are shown: 1) before the eclipse
    (black solid lines); 2) during the first half of the eclipse
    (blueshifted red dashed lines); 3) during eclipse maximum (blue
    dashed lines); and 4) during the second half of the eclipse
    (redshifted red dashed lines). The configurations of the four
    phases are shown in Fig.\,\ref{fig:Eclipse}.}
\end{figure}

With our eclipse observations we can test the hypothesis that differences
between synthetic and observed spectra are more severe in smaller regions of
the solar disk. We show a comparison of modeled and observed spectra during
different phases of the eclipse in Fig.\,\ref{fig:SpecComparison1}. For each
case, we show four different spectra: 1) a spectrum out of eclipse; 2) a
spectrum taken at a phase when about 60\,\% of the disk is occulted by the
Moon, most of the visible part is on the blueshifted side of the Sun
(spectrum taken at 09:24\,UT, see upper right panel in
Fig.\,\ref{fig:Eclipse}); 3) a spectrum taken at eclipse maximum, the visible
is almost symmetrically distributed between the blue- and redshifted parts of
the disk (09:42\,UT, lower left panel in Fig.\,\ref{fig:Eclipse}); and 4) a
spectrum like in 2), but after eclipse maximum and most of the visible disk
red-shifted (10:00\,UT, bottom middle panel in Fig.\,\ref{fig:Eclipse}).

The comparison in Fig.\,\ref{fig:SpecComparison1} shows that out of
eclipse and during eclipse maximum, the observed and modeled spectra
compare quite well. The spectrum taken at eclipse maximum has a
shallower core and wider wings than the spectrum taken out of
eclipse. This behavior is well reproduced by our model. The two
spectra observed during the first and second half of the eclipse (red
dashed lines) are blue- and redshifted, respectively, and they both
show a line core depth between the spectra taken out of eclipse and at
eclipse maximum. In contrast to this, the modeled line profiles at
these phases are significantly deeper. There is almost no difference
between the line depth of these profiles and the spectrum modeled out
of eclipse. This is a clear discrepancy to our observations that we
ascribe to incorrectly modeled line depths, in particular at
intermediate $\mu$-values ($\mu > 0.6$), because at eclipse maximum
mostly regions with lower $\mu$-values are visible. Here, the
calculated line depth seems to reproduce the observations quite well.

\begin{figure}
  \centering
  \resizebox{\hsize}{!}{\includegraphics[bb = 30 10 625 445]{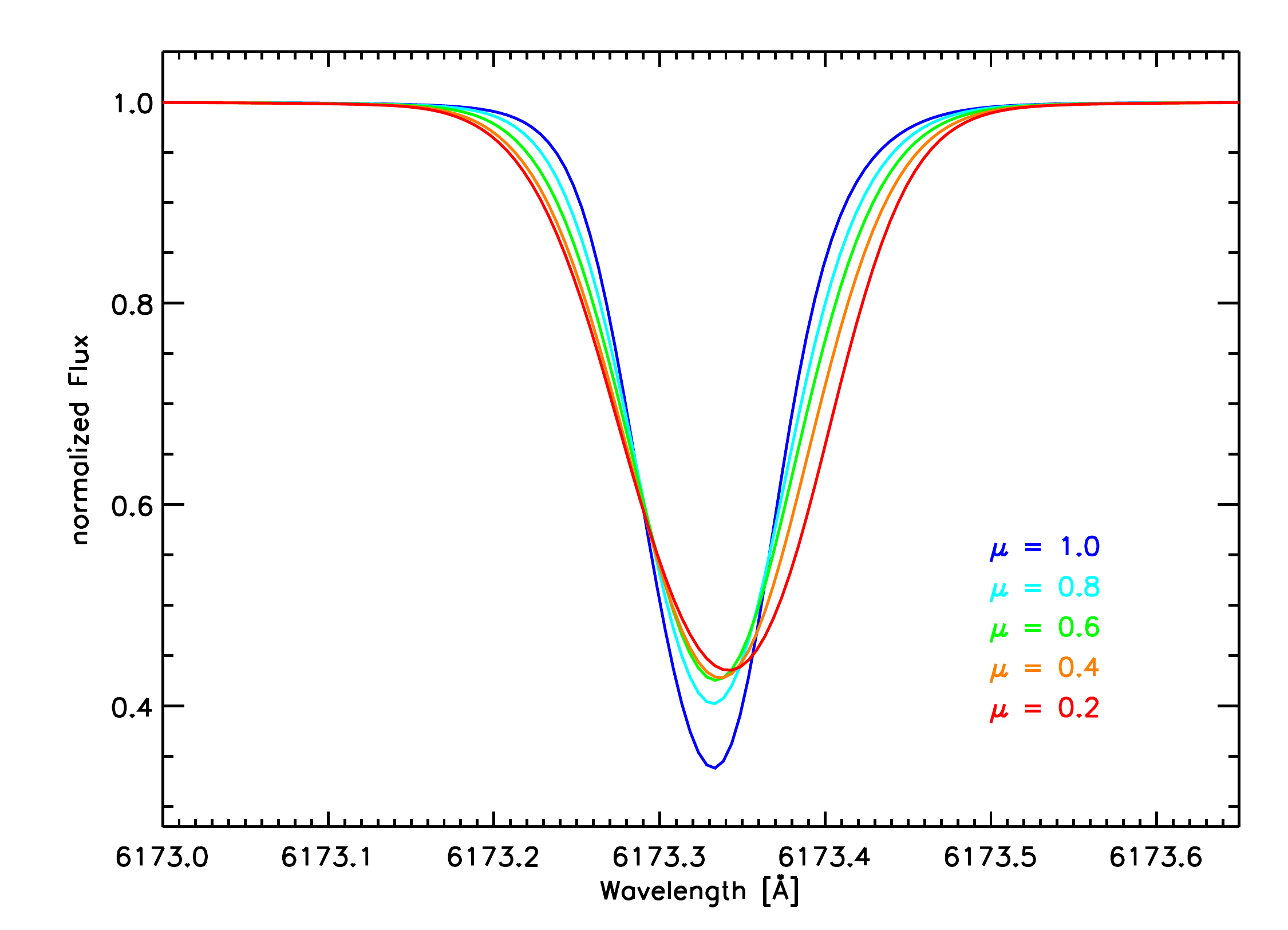}}
  \caption{\label{fig:SpecComparison2}Synthetic line profiles for
    solar surface areas at different $\mu$-values from
    \citet{2013A&A...558A..49B}.}
\end{figure}

To shed some light on the question whether incorrectly modeled line depths at
intermediate $\mu$-values can be the reason for the discrepancy between our
observed eclipse spectra and our model, we show the synthetic line profiles
from \cite{2013A&A...558A..49B} in Fig.\,\ref{fig:SpecComparison2}. The strong
blueshift of the line at $\mu = 1$ is clearly visible relative to the other
line profiles. Furthermore, the $\mu = 1$ profile is rather narrow but much
deeper than all other profiles, and the line depths show a steep gradient from
$\mu =1$ to $\mu = 0.6$. A detailed investigation of possible uncertainties in
the modeled line profiles is beyond the scope of this paper. Nevertheless,
we note that the fast change of line depth at limb angles between $\mu = 0.6$
and $\mu = 1$ could imply that line depths for individual limb angles may be
somewhat uncertain, or that our interpolation between the calculated profiles
is incorrect. In any case, a detailed comparison between model spectra and
observations at well-defined limb angles, in particular between $\mu = 0.6$
and $\mu = 1$, would be highly desirable.

\section{Summary}

On March 20, 2015, we obtained 159 high-resolution FTS spectra of the
Sun as a star, 76 of them during partial solar eclipse. From each
spectrum, we derived the apparent absolute RV. Before and after the
eclipse, our RVs are tracking barycentric motion. During eclipse, they
provide an exquisite benchmark RM curve with a peak-to-peak amplitude
of almost 1.4\,km\,s$^{-1}$. The geometry of the event is fully
determined by our knowledge about solar, lunar, and terrestrial
motion. Doppler velocities on the solar surface and limb darkening are
also well known, but information about the limb angle
dependence of solar convective blueshift and spectral line shape
is incomplete. With
this paper, we publish the RV values of our observations. The time
series can be used to test line profile calculations and codes
calculating RM curves for exoplanet observations. The amplitude of the
RM curve is orders of magnitudes higher than in typical exoplanet
observations, while the uncertainties in the RVs are very small
(S/N$>1000$). In contrast to other stars, much information is
available about solar surface velocities.

We tested whether we were able to reproduce the RM curve during solar eclipse
using a set of different approaches. First, we parameterized limb
darkening and solar surface rotation using standard descriptions but
neglected convective blueshift. We calculated surface integrated RVs
by computing the intensity-weighted average of the radial velocities
over the visible solar disk. This first attempt resulted in an RV
curve that qualitatively matched the observations, but revealed a clear
offset with the RM curve being off to the blue. In our second model,
we included a qualitative prescription of convective blueshift leaving
all other steps as in our first model. This resulted in a much better
reproduction of the observed RV curve, but still revealed residuals as
high as 40\,m\,s$^{-1}$ especially during the first half of the
eclipse. In our model~3 we calculated spectral line profiles for
different locations on the solar surface and computed a
disk-integrated line from which we determined the apparent solar RVs
like in our observations. For the line profiles, we used results from
a 3D-MHD calculation that inherently includes information on
convective blueshift and the variation of the profile shape. While
this model provided the best match to our observations, residuals were
still as high as 30\,m\,s$^{-1}$, the relative errors were smaller
than 10\,\%. As a consistency check, we used solar
surface observations from HMI in model 4 to describe limb darkening and solar
velocity fields. The results of this exercise are very similar to
those from model~1. This showed the limitation of the HMI
Dopplergrams for full solar disk calculations. It is
relevant for investigations of convective blueshift that use
Dopplergram observations as reference \citep[e.g.,][using SOHO/MDI and
SDO/HMI, respectively]{2010A&A...519A..66M, 2016MNRAS.457.3637H}.

With the goal to identify the reason for the mismatch between our
models and the RV observations, we inspected the line profiles modeled
and observed during different phases of the eclipse. We found a
mismatch of line depths close to phases 0.25 and 0.75, but a rather
good match at phase 0.5 and outside eclipse. We concluded that the
line depths at intermediate limb angles, $\mu$ between 0.6 and 1, may
be systematically overestimated, which could lead to an incorrect
weighting of velocities across the solar disk. Identifying reasons for
this problem is beyond the scope of this work, but we see the need
to obtain high-quality spectral observations of the solar surface at
well-defined limb angles with high wavelength accuracy.

As the main cause for the mismatch between our model calculations and observed
RVs, we identified our limited knowledge about line profiles and convective
blueshift as a function of solar limb angle. Alternative explanations include
the variation of blueshift with magnetic field strength, incorrect assumptions
about solar rotation and granulation, and so far undiscovered mechanisms
affecting the spectral observation of the spatially extended solar disk during
eclipse. The solution of this puzzle will have ramifications for solar physics
because a better understanding of the local line profiles is required, and for
exoplanet research because it will lead to a more accurate description of the
RM curve in particular, and of stellar RVs in general.

\begin{acknowledgements}
  We thank Jasper Schou, Achim Gandorfer, and Mathias Zechmeister for helpful
  discussions, and the team at IAG for their support with the
  observations. Part of this work was supported by the ERC Starting Grant
  \emph{Wavelength Standards}, Grant Agreement Number 279347. AR acknowledges
  research funding from DFG grant RE 1664/9-1. The FTS was funded by the DFG
  and the State of Lower Saxony through the Gro{\ss}ger\"ateprogramm
  \emph{Fourier Transform Spectrograph}.
\end{acknowledgements}

\bibliographystyle{aa}
\bibliography{refs}


\clearpage
\onecolumn

  \begin{longtable}{rcccrc}
    \caption{\label{tab:Observations}Log of observations}\\
    \hline\hline
    \# & UT &  Airmass & Ecl.\ Frac. & RV\,\,\,\, & SNR \\
    &&& [$\%$] &  [m\,s$^{-1}$] & \\
    \hline
    \endfirsthead
    \caption{continued}\\
    \hline\hline
    \# & UT &  Airmass & Ecl.\ Frac. & RV\,\,\,\,  & SNR \\
    &&& [$\%$] &  [m\,s$^{-1}$] & \\
    \hline
    \endhead
    \hline
    \endfoot
      1 &  7:07:57.2 & 3.79 &  0.0 &  160.4 & 170 \\
  2 &  7:09:45.8 & 3.73 &  0.0 &  160.1 & 170 \\
  3 &  7:11:34.3 & 3.67 &  0.0 &  161.5 & 170 \\
  4 &  7:13:22.9 & 3.61 &  0.0 &  161.3 & 170 \\
  5 &  7:15:11.5 & 3.56 &  0.0 &  164.0 & 190 \\
  6 &  7:17:00.3 & 3.50 &  0.0 &  162.8 & 180 \\
  7 &  7:18:49.7 & 3.45 &  0.0 &  166.8 & 160 \\
  8 &  7:20:38.5 & 3.40 &  0.0 &  168.6 & 160 \\
  9 &  7:22:27.4 & 3.35 &  0.0 &  169.9 & 170 \\
 10 &  7:24:16.5 & 3.30 &  0.0 &  167.3 & 170 \\
 11 &  7:26:05.5 & 3.26 &  0.0 &  168.6 & 180 \\
 12 &  7:27:53.8 & 3.21 &  0.0 &  168.9 & 190 \\
 13 &  7:29:42.5 & 3.17 &  0.0 &  168.0 & 180 \\
 14 &  7:31:30.7 & 3.13 &  0.0 &  173.1 & 190 \\
 15 &  7:33:19.2 & 3.09 &  0.0 &  169.1 & 170 \\
 16 &  7:35:09.5 & 3.05 &  0.0 &  166.9 & 190 \\
 17 &  7:36:58.0 & 3.01 &  0.0 &  172.8 & 200 \\
 18 &  7:38:46.1 & 2.98 &  0.0 &  170.1 & 180 \\
 19 &  7:40:34.6 & 2.94 &  0.0 &  168.4 & 170 \\
 20 &  7:42:22.8 & 2.91 &  0.0 &  172.3 & 170 \\
 21 &  7:44:11.5 & 2.87 &  0.0 &  172.3 & 190 \\
 22 &  7:46:00.1 & 2.84 &  0.0 &  170.0 & 180 \\
 23 &  7:47:48.8 & 2.81 &  0.0 &  177.3 & 190 \\
 24 &  7:49:37.1 & 2.78 &  0.0 &  175.8 & 180 \\
 25 &  7:51:25.3 & 2.75 &  0.0 &  175.5 & 180 \\
 26 &  7:53:13.9 & 2.72 &  0.0 &  170.8 & 180 \\
 27 &  7:55:02.3 & 2.69 &  0.0 &  174.3 & 190 \\
 28 &  7:56:50.7 & 2.66 &  0.0 &  179.9 & 180 \\
 29 &  7:58:39.0 & 2.63 &  0.0 &  177.8 & 180 \\
 30 &  8:00:27.2 & 2.61 &  0.0 &  176.5 & 180 \\
 31 &  8:02:15.7 & 2.58 &  0.0 &  179.3 & 200 \\
 32 &  8:04:04.0 & 2.56 &  0.0 &  180.3 & 210 \\
 33 &  8:05:53.0 & 2.53 &  0.0 &  182.3 & 200 \\
 34 &  8:07:41.4 & 2.51 &  0.0 &  181.5 & 180 \\
 35 &  8:09:30.0 & 2.49 &  0.0 &  184.4 & 190 \\
 36 &  8:11:18.5 & 2.46 &  0.0 &  182.8 & 190 \\
 37 &  8:13:07.1 & 2.44 &  0.0 &  184.6 & 170 \\
 38 &  8:14:55.3 & 2.42 &  0.0 &  184.2 & 190 \\
 39 &  8:16:43.7 & 2.40 &  0.0 &  183.7 & 190 \\
 40 &  8:18:32.1 & 2.38 &  0.0 &  185.2 & 200 \\
 41 &  8:20:20.4 & 2.36 &  0.0 &  186.8 & 210 \\
 42 &  8:22:09.4 & 2.34 &  0.0 &  187.2 & 200 \\
 43 &  8:23:57.7 & 2.32 &  0.0 &  189.4 & 190 \\
 44 &  8:25:46.1 & 2.30 &  0.0 &  193.1 & 210 \\
 45 &  8:27:34.7 & 2.28 &  0.0 &  192.6 & 190 \\
 46 &  8:29:23.0 & 2.26 &  0.0 &  191.9 & 200 \\
 47 &  8:31:11.3 & 2.25 &  0.0 &  190.5 & 190 \\
 48 &  8:32:59.5 & 2.23 &  0.0 &  196.5 & 180 \\
 49 &  8:34:47.9 & 2.21 &  0.2 &  189.6 & 200 \\
 50 &  8:36:36.3 & 2.20 &  1.0 &  180.7 & 210 \\
 51 &  8:38:54.5 & 2.18 &  2.5 &  172.2 & 210 \\
 52 &  8:40:43.0 & 2.16 &  3.9 &  157.0 & 200 \\
 53 &  8:42:31.4 & 2.15 &  5.5 &  139.4 & 190 \\
 54 &  8:44:19.9 & 2.13 &  7.2 &  121.0 & 200 \\
 55 &  8:46:08.5 & 2.12 &  9.1 &  102.4 & 200 \\
 56 &  8:47:56.8 & 2.10 & 11.0 &   80.8 & 190 \\
 57 &  8:49:45.5 & 2.09 & 13.1 &   61.1 & 200 \\
 58 &  8:51:34.0 & 2.08 & 15.3 &   40.0 & 220 \\
 59 &  8:53:22.9 & 2.06 & 17.6 &   14.2 & 220 \\
 60 &  8:55:11.5 & 2.05 & 19.9 &  -12.7 & 190 \\
 61 &  8:56:59.6 & 2.04 & 22.3 &  -37.9 & 200 \\
 62 &  8:58:47.9 & 2.03 & 24.7 &  -62.6 & 180 \\
 63 &  9:00:36.1 & 2.01 & 27.2 &  -89.6 & 200 \\
 64 &  9:02:24.8 & 2.00 & 29.8 & -117.7 & 200 \\
 65 &  9:04:13.1 & 1.99 & 32.4 & -142.7 & 200 \\
 66 &  9:06:01.7 & 1.98 & 35.1 & -168.9 & 180 \\
 67 &  9:07:50.2 & 1.97 & 37.7 & -196.9 & 190 \\
 68 &  9:09:38.7 & 1.96 & 40.4 & -219.2 & 180 \\
 69 &  9:11:27.1 & 1.95 & 43.1 & -245.6 & 170 \\
 70 &  9:13:15.5 & 1.94 & 45.7 & -269.7 & 180 \\
 71 &  9:15:04.0 & 1.93 & 48.4 & -283.7 & 180 \\
 72 &  9:16:53.1 & 1.92 & 51.1 & -302.9 & 170 \\
 73 &  9:18:42.1 & 1.91 & 53.7 & -317.9 & 190 \\
 74 &  9:20:30.9 & 1.90 & 56.3 & -324.8 & 170 \\
 75 &  9:22:19.7 & 1.89 & 58.9 & -325.9 & 170 \\
 76 &  9:24:08.5 & 1.88 & 61.3 & -322.9 & 160 \\
 77 &  9:25:57.3 & 1.87 & 63.7 & -311.9 & 150 \\
 78 &  9:27:45.9 & 1.86 & 66.0 & -283.6 & 150 \\
 79 &  9:29:34.7 & 1.85 & 68.1 & -251.3 & 150 \\
 80 &  9:31:23.1 & 1.85 & 70.0 & -203.1 & 140 \\
 81 &  9:33:11.5 & 1.84 & 71.7 & -138.2 & 140 \\
 82 &  9:34:59.9 & 1.83 & 73.2 &  -53.8 & 140 \\
 83 &  9:36:48.4 & 1.82 & 74.4 &   42.8 & 140 \\
 84 &  9:38:37.4 & 1.81 & 75.2 &  157.3 & 140 \\
 85 &  9:40:26.3 & 1.81 & 75.7 &  280.3 & 130 \\
 86 &  9:42:15.0 & 1.80 & 75.9 &  398.3 & 140 \\
 87 &  9:44:03.8 & 1.79 & 75.6 &  519.5 & 140 \\
 88 &  9:45:52.2 & 1.79 & 75.0 &  631.8 & 140 \\
 89 &  9:47:40.7 & 1.78 & 74.0 &  725.3 & 140 \\
 90 &  9:49:29.7 & 1.77 & 72.8 &  809.2 & 140 \\
 91 &  9:51:18.4 & 1.77 & 71.2 &  873.3 & 130 \\
 92 &  9:53:07.5 & 1.76 & 69.5 &  919.9 & 140 \\
 93 &  9:54:55.9 & 1.75 & 67.5 &  955.7 & 150 \\
 94 &  9:56:44.9 & 1.75 & 65.5 &  977.1 & 160 \\
 95 &  9:58:33.3 & 1.74 & 63.2 &  986.6 & 170 \\
 96 & 10:00:21.8 & 1.74 & 60.9 &  985.0 & 170 \\
 97 & 10:02:10.1 & 1.73 & 58.5 &  980.4 & 180 \\
 98 & 10:03:58.6 & 1.73 & 56.0 &  966.0 & 160 \\
 99 & 10:05:47.4 & 1.72 & 53.4 &  952.1 & 180 \\
100 & 10:07:36.2 & 1.72 & 50.9 &  930.8 & 160 \\
101 & 10:09:54.5 & 1.71 & 47.6 &  895.0 & 190 \\
102 & 10:11:43.9 & 1.71 & 45.0 &  874.6 & 190 \\
103 & 10:13:33.6 & 1.70 & 42.4 &  842.9 & 170 \\
104 & 10:15:22.6 & 1.70 & 39.8 &  814.1 & 190 \\
105 & 10:17:11.7 & 1.69 & 37.3 &  787.0 & 190 \\
106 & 10:19:00.9 & 1.69 & 34.7 &  756.6 & 190 \\
107 & 10:20:49.9 & 1.68 & 32.2 &  723.6 & 180 \\
108 & 10:22:38.9 & 1.68 & 29.7 &  693.3 & 210 \\
109 & 10:24:27.9 & 1.68 & 27.2 &  662.2 & 210 \\
110 & 10:26:17.0 & 1.67 & 24.9 &  631.3 & 220 \\
111 & 10:28:07.1 & 1.67 & 22.5 &  602.6 & 210 \\
112 & 10:29:56.1 & 1.66 & 20.2 &  572.5 & 210 \\
113 & 10:31:45.1 & 1.66 & 18.0 &  545.8 & 220 \\
114 & 10:33:34.0 & 1.66 & 15.8 &  515.3 & 200 \\
115 & 10:35:22.9 & 1.65 & 13.8 &  488.8 & 210 \\
116 & 10:37:12.0 & 1.65 & 11.8 &  467.3 & 220 \\
117 & 10:39:01.0 & 1.65 &  9.9 &  442.9 & 220 \\
118 & 10:40:49.9 & 1.65 &  8.1 &  421.3 & 210 \\
119 & 10:42:38.7 & 1.64 &  6.4 &  401.7 & 200 \\
120 & 10:44:27.6 & 1.64 &  4.9 &  383.3 & 220 \\
121 & 10:46:16.8 & 1.64 &  3.4 &  367.8 & 210 \\
122 & 10:48:05.8 & 1.64 &  2.2 &  351.2 & 200 \\
123 & 10:49:54.9 & 1.63 &  1.1 &  348.8 & 210 \\
124 & 10:51:43.6 & 1.63 &  0.3 &  345.7 & 200 \\
125 & 10:53:32.7 & 1.63 &  0.0 &  346.7 & 210 \\
126 & 10:55:21.9 & 1.63 &  0.0 &  349.9 & 220 \\
127 & 10:57:10.8 & 1.63 &  0.0 &  353.9 & 230 \\
128 & 10:58:59.7 & 1.62 &  0.0 &  359.1 & 200 \\
129 & 11:00:49.8 & 1.62 &  0.0 &  357.1 & 220 \\
130 & 11:02:38.6 & 1.62 &  0.0 &  361.5 & 220 \\
131 & 11:04:27.7 & 1.62 &  0.0 &  358.9 & 200 \\
132 & 11:06:16.8 & 1.62 &  0.0 &  362.8 & 220 \\
133 & 11:08:05.7 & 1.62 &  0.0 &  363.9 & 230 \\
134 & 11:09:54.6 & 1.62 &  0.0 &  367.3 & 190 \\
135 & 11:11:43.6 & 1.62 &  0.0 &  369.7 & 210 \\
136 & 11:13:33.1 & 1.62 &  0.0 &  372.9 & 200 \\
137 & 11:15:22.0 & 1.61 &  0.0 &  373.3 & 210 \\
138 & 11:17:10.9 & 1.61 &  0.0 &  380.0 & 200 \\
139 & 11:18:59.9 & 1.61 &  0.0 &  380.8 & 210 \\
140 & 11:20:48.7 & 1.61 &  0.0 &  382.8 & 220 \\
141 & 11:22:37.7 & 1.61 &  0.0 &  382.4 & 200 \\
142 & 11:24:26.7 & 1.61 &  0.0 &  385.8 & 200 \\
143 & 11:26:15.7 & 1.61 &  0.0 &  387.8 & 200 \\
144 & 11:28:04.3 & 1.61 &  0.0 &  393.2 & 200 \\
145 & 11:29:53.0 & 1.61 &  0.0 &  395.6 & 220 \\
146 & 11:31:41.7 & 1.61 &  0.0 &  398.7 & 210 \\
147 & 11:33:30.3 & 1.61 &  0.0 &  400.4 & 210 \\
148 & 11:35:19.0 & 1.61 &  0.0 &  400.3 & 200 \\
149 & 11:37:07.7 & 1.61 &  0.0 &  407.5 & 210 \\
150 & 11:38:56.9 & 1.61 &  0.0 &  408.3 & 210 \\
151 & 11:48:37.5 & 1.62 &  0.0 &  413.3 & 230 \\
152 & 11:50:26.6 & 1.62 &  0.0 &  416.8 & 200 \\
153 & 11:52:15.6 & 1.62 &  0.0 &  424.7 & 220 \\
154 & 11:54:04.2 & 1.62 &  0.0 &  422.9 & 200 \\
155 & 11:55:53.1 & 1.62 &  0.0 &  426.1 & 210 \\
156 & 11:57:41.8 & 1.62 &  0.0 &  426.1 & 210 \\
157 & 11:59:30.5 & 1.63 &  0.0 &  431.1 & 210 \\
158 & 12:01:19.1 & 1.63 &  0.0 &  434.5 & 210 \\
159 & 12:03:07.6 & 1.63 &  0.0 &  434.0 & 230 \\

  \end{longtable}


\end{document}